\documentclass[letterpaper,11pt]{article}
\usepackage{amsmath, amsthm, amssymb}
\usepackage[usenames, dvipsnames]{color}
\usepackage[normalem]{ulem} 
\usepackage{fullpage}
\usepackage[numbers, sort]{natbib}
\usepackage[noend]{algorithmic}
\usepackage{tikz, pgfplots}
\usepackage{enumerate}
\usepackage{hyperref}
\usepackage{xspace,color}
\usepackage{graphicx}
\usepackage{caption}
\usepackage{subcaption}
\usepackage{enumitem,linegoal}
\usepackage[linesnumbered,ruled,vlined]{algorithm2e}
\usepackage{comment}	
\usepackage{ctable}	
\usepackage{mathtools}
\usepackage[flushleft]{threeparttable}
\usepackage{verbatim}
\usepackage{setspace}
\usepackage{bbm}
\usepackage{thm-restate}
\usepackage{footnote}
\usepackage{tcolorbox}
\usepackage{fdsymbol}

\usepackage{apxproof}

\makesavenoteenv{tabular}
\makesavenoteenv{table}

\usepackage[margin=1in]{geometry}

\hypersetup{
	colorlinks   = true, 
	urlcolor     = blue, 
	linkcolor    = blue, 
	citecolor   = red 
}

\definecolor{crimsonglory}{rgb}{0,0,0}

\newtheorem{example}{Example}

\newtheorem{theorem}{Theorem}[section]

\newtheorem{lemma}[theorem]{Lemma}

\newtheorem{corollary}[theorem]{Corollary}

\newtheorem{definition}[theorem]{Definition}

\newtheorem{observation}{Observation}[section]

\newcommand{\etal}{\textit{et al. }}

\newtheorem{remark}[theorem]{Remark}
\makeatletter
\def\GrabProofArgument[#1]{ #1: \egroup\ignorespaces}
\def\proof{\noindent\textbf\bgroup Proof%
	\@ifnextchar[{\GrabProofArgument}{. \egroup\ignorespaces}}

\makeatother

\usetikzlibrary{shapes.geometric, arrows.meta, positioning}

\tikzstyle{startstop} = [rectangle, rounded corners, minimum width=3cm, minimum height=1cm, text centered, draw=black, fill=red!30]
\tikzstyle{process} = [rectangle, minimum width=3cm, minimum height=1cm, text centered, draw=black, fill=orange!30]
\tikzstyle{decision} = [diamond, minimum width=3cm, minimum height=1cm, text centered, draw=black, fill=green!30]
\tikzstyle{arrow} = [thick,->,>=stealth]

\usepackage[latin1]{inputenc}

\newcommand{\efx}{\mathsf{EFX}}
\newcommand{\eftx}{\mathsf{EF2X}}
\newcommand{\kefx}[1]{#1\mathsf{-EFX}}
\newcommand{\nsw}{\mathsf{NSW}}
\newcommand{\valu}{v}

\newcommand{\invsq}{\frac{\sqrt{2}}{2}}

\newcommand{\goods}{M}

\newcommand{\sgoods}{S}
\newcommand{\agents}{N}

\newcommand{\sagents}{S}
\newcommand{\alloc}{X}
\DeclareRobustCommand{\rchi}{{\mathpalette\irchi\relax}}
\newcommand{\irchi}[2]{\raisebox{\depth}{$#1\chi$}} 
\newcommand{\malloc}{\rchi}

\newcommand{\comp}{\mathcal{C}}

\newcommand{\rank}{\mathfrak{R}}
\newcommand{\rupp}{\mathfrak{r}}
\newcommand{\virtualvalue}{\mathfrak v}
\newcommand{\instance}{\mathcal I}
\newcommand{\values}{\mathcal V}
\newcommand{\algo}{\textsf{CXXRA}}
\newcommand{\algot}{\textsf{C$\sqrt{2}$XRA}}
\newcommand{\algoth}{\textsf{C$\sqrt{2}$XPQ}}
\newcommand{\algopr}{\textsf{PQRAX}}

\newcommand{\good}{g}

\newcommand{\relevant}[2]{#1{[#2]}}
\newcommand{\roott}[2]{\textsf{root}({#1},{#2})}
\newcommand{\cycle}{C}

\newtheorem{theoremrep}{Theorem}[section]
\newtheorem{lemmarep}[theorem]{Lemma}
\newtheorem{observationrep}{Observation}[section]

\newcounter{proccnt}

\newcommand{\konote}[1]{}

\usepackage[normalem]{ulem} 

\title{Almost Envy-free Allocation of Indivisible Goods:\\ A Tale of Two Valuations}

\author{
	Alireza Kaviani\thanks{Sharif University of Technology, Tehran, Iran}
	\and
	Masoud Seddighin\thanks{Tehran Institute for Advanced Studies (TeIAS), Tehran, Iran}
	\and
	AmirMohammad Shahrezaei\footnotemark[1]
}

\SetCommentSty{mycommfont}

\SetKwInput{KwInput}{Input}                
\SetKwInput{KwOutput}{Output}  
\begin{document}
	\newcommand{\ignore}[1]{}
	\renewcommand{\theenumi}{\roman{enumi}.}
	\renewcommand{\labelenumi}{\theenumi}
	\sloppy
	\date{} 
	\newenvironment{subproof}[1][\proofname]{
		\renewcommand{\Box}{ \blacksquare}%
		\begin{proof}[#1]%
		}{%
		\end{proof}%
	}
	\maketitle
	
	\thispagestyle{empty}
	\allowdisplaybreaks

	\vspace{0.5cm}
	\begin{abstract}
		The existence of $\efx$ allocations stands as one of the main challenges in discrete fair division.In this paper, we present symmetrical results on the existence of $\efx$ and its approximate variations for two distinct valuations: restricted additive valuations and $(p,q)$-bounded valuations introduced by Christodoulou \etal \cite{christodoulou2023fair}. In a $(p,q)$-bounded instance, each good has relevance for at most $p$ agents, and any pair of agents shares at most $q$ common relevant goods.

We show that instances with $(\infty,1)$-bounded valuations admit $\eftx$ allocations and $\efx$ allocations with at most $\lfloor {n}/{2} \rfloor - 1$ discarded goods, mirroring results for the restricted additive setting \cite{akrami2022ef2x}. We also present $\kefx{({\sqrt{2}}/{2})}$ algorithms for both restricted additive and $(\infty,1)$-bounded subadditive settings. The symmetry of these results suggests these valuations share symmetric structures. Building on this, we propose an $\efx$ allocation for restricted additive valuations when $p=2$ and $q=\infty$.

To achieve these results, we further develop the rank concept introduced by Farhadi \etal \cite{farhadi2021almost} and introduce several new concepts such as virtual value, rankpath, and root, which advance the overall understanding of $\efx$ allocations. In addition, we suggest an updating rule based on the virtual values  which we believe will lead to broader and more generalized results on $\efx$.

	\end{abstract}

\section{Introduction}\label{introduction}


Fair division is a longstanding problem in mathematics and economics \cite{Steinhaus:first,brams,brams1996fair,even1984note,stromquist1980cut,Dubins:first}. It seeks methods to allocate resources so that each party receives a fair share. Modern research on this problem began with Steinhaus in 1945, by introducing the cake-cutting problem \cite{Steinhaus:first}. This problem involves dividing a heterogeneous, divisible resource among agents with different preferences, such that each individual believes her share is fair. Since the introduction of cake-cutting, numerous fairness notions have emerged, such as proportionality, envy-freeness, and equity. Fortunately, scientists have significantly advanced in providing guarantees for these criteria; see \cite{brams1996fair} for an overview.


Over the past two decades, a discrete version of the cake-cutting problem, known as fair allocation, has garnered interest among computer scientists \cite{Budish:first,Bezacova:first,Procaccia:first,Saberi:first,ghodsi2018fair,barman2017finding,caragiannis2016unreasonable,akrami2023efx}. In the fair allocation problem, the resource is a set of indivisible goods rather than a divisible cake.
Fair allocation of indivisible goods presents many intriguing unresolved questions. In this paper, we investigate one of the most challenging: envy-freeness up to any good, or $\efx$ --- referred to by Ariel Procaccia as "\textit{Fair Division's Most Enigmatic Question}" \cite{procaccia2020technical}. Consider a set $\goods$ of indivisible goods and $n$ agents. Each agent $i$ has a monotone valuation function $\valu_i: 2^\goods \rightarrow \mathbb{R}^{\geq 0}$ that assigns a non-negative value to each subset of goods. If we allocate bundle $\alloc_i$ to agent $i$ for $1 \leq i \leq n$, the allocation is $\efx$ if, for every $1 \leq i, j \leq n$ and every ${\alloc'_j \subset \alloc_j}$ we have $\valu_i(\alloc_i) \ge \valu_i(\alloc'_j).
$

$\efx$ is indeed, a relaxation of the more stringent notion called \textit{``envy-freeness"}. In an envy-free allocation, each agent prefers their bundle over any other agent's bundle, meaning $\valu_i(\alloc_i) \ge \valu_i(\alloc'_j)$ must hold even when $\alloc'_j = \alloc_j$. While examples easily disprove the existence of envy-free allocations, it remains an open question whether $\efx$ allocations are always possible.
As of now, $\efx$ allocations are known to exist only in a few limited scenarios. For instance, with two agents having monotone valuations, an $\efx$ allocation can always be found \cite{plaut2018almost}. A similar result holds for three agents where two have monotone valuations, and one has an \emph{MMS-feasible} valuation \cite{akrami2023efx,chaudhury2020efx}. There are also some guarantees proved for an arbitrary number of agents. For additive valuations, we can always guarantee $0.618$-$\efx$ \cite{amanatidis2020multiple,farhadi2021almost}. \footnote{See Section \ref{sec:bc} for a formal definition of $\beta$-$\efx$.} Moreover, for monotone subadditive valuations, an $\efx$ allocation exists that discards at most $n-1$ goods \cite{chaudhury2021little}. For a detailed overview, refer to Section \ref{sec:related}.

In recent advances related to $\efx$ allocations, our focus has been drawn to two separate studies that provide $\efx$ guarantees in distinct scenarios. The first study, conducted by Christodoulou \etal \cite{christodoulou2023fair}, introduces a family of valuation functions where each good holds relevance (i.e., has a non-zero marginal value) for at most $p$ agents, and any pair of agents share at most $q$ common relevant goods. There are several real-world motivations for this setting. 
Examples include allocating office space in academic settings, disaster response planning, and delineating ``areas of influence" among neighboring powers \cite{christodoulou2023fair}.
They present an $\efx$ allocation for the case where $p=2$ and $q=1$, leaving the more general cases as open directions:
\begin{quote}
	"\textit{Unless one can solve the basic question if $\efx$ allocations always exist -- partial progress can possibly be made for larger values of $p$ and $q$}" \cite{christodoulou2023fair}.
\end{quote}

In the second study, Akrami \etal \cite{akrami2023efx} consider $\efx$ allocations for the restricted additive valuations, where the valuations are additive, but the value of every good $\good$ for every agent is restricted to be either $0$ or $\valu_\good$. 
This setting has been extensively studied for different allocation problems, particularly for the maximin fairness notion, which is commonly known as the Santa Claus problem. For a detailed discussion on the significance and applications of the restricted additive valuations, refer to \cite{akrami2022ef2x}.
Akrami \etal \cite{akrami2022ef2x} prove the existence of $\eftx$ allocations for the restricted additive setting. An allocation $\langle\alloc_1,\alloc_2,\ldots,\alloc_n\rangle$ is $\eftx$, if for every agent $i$ and $j$ and every  
$\alloc'_j \subset \alloc_j$ such that $|\alloc'_j| \leq |\alloc_j|-2$, we have $\valu_i(\alloc_i)\ge \valu_i(\alloc'_j). $

In this paper, we uncover intriguing parallels between these two seemingly different settings --- the restricted additive setting and $(p,q)$-bounded valuations (Table \ref{tab:my-table}). 
The results in this paper, along with those from \cite{christodoulou2023fair} and \cite{akrami2022ef2x}, form a symmetric set of results: 

\begin{quote}
when $p$ is either $2$ or $\infty$, every allocation algorithm with $\efx$, $\eftx$, or approximate $\efx$ guarantees obtained for $(p,1)$-bounded valuations has a corresponding algorithm with the same guarantee for $(p,\infty)$-bounded valuations with the restricted additive property. 
\end{quote}

The algorithms we use in our work introduce several techniques and ideas that may be of independent interest. In particular, we revisit the concept of rank introduced by Farhadi \etal \cite{farhadi2021almost} and, based on this concept, introduce the virtual welfare of the agents for an allocation, which may be of independent interest.
You can find a detailed explanation of these techniques in Section \ref{sec:resultstechniques}. 
These results can serve as a building block for providing guarantees for larger values of $p$ and $q$. We note that graphical models like $(p,q)$-bounded valuations are commonly used as a middle ground to pave the way toward proving more general results. This approach has already been used to achieve strong results in algorithmic game theory, such as the complexity of Nash equilibrium \cite{chen2009settling,daskalakis2009complexity}. 
	\section{Basic Concepts} \label{sec:bc}
Every allocation problem instance is represented as a triple $\instance = (\agents,\goods,\values)$, where $\agents$ is a set of $n$ agents, $\goods$ is a set of $m$ goods, and $\values$ is the set of valuation functions. We refer to agents by their index from $1$ to $n$ and use the variable $\good$ to refer to a good. Our results in this paper relate to two types of valuation functions. Let us first define them formally.  
\begin{definition}
	Let $\mathcal V = \{\valu_1,\valu_2,\ldots,\valu_n\}$ be a set of additive valuation functions where for every $i$,  $\valu_i: 2^\goods \rightarrow R^{\geq 0}$. We say $\mathcal{V}$ is restricted additive, if for every good $\good$ and every $1 \leq i \leq n$,	we have $\valu_i(\{\good\}) \in \{0,\valu_\good\}$. For convenience, for every subset $S$ of goods, we define $\valu(S) = \sum_{g\in S} \valu_g$.
\end{definition}

The second type of valuations we study is $(p,q)$-bounded valuations. 

\begin{definition}\label{def1}[Christodoulou \etal ~\cite{christodoulou2023fair}]
Let $\mathcal{V} = \{\valu_1, \valu_2, \ldots, \valu_n\}$ be a set of monotone valuation functions, where each $\valu_i: 2^\goods \rightarrow \mathbb{R}^{\geq 0}$. A good $\good$ is not relevant to agent $i$ if
$
\forall \alloc_i \subseteq \goods \setminus {\good}, \valu_i(\alloc_i \cup {\good}) = \valu_i(\alloc_i),
$
and relevant otherwise. Set $\mathcal{V}$ is $(p,q)$-bounded if each good is relevant to at most $p$ agents, and there are at most $q$ goods relevant to any pair of agents.
\end{definition}
Note that the valuations in $\mathcal{V}$ are not necessarily additive; we only require them to be monotone. Here, we consider a slightly adjusted variant with an additional assumption that the valuations are strictly monotone for relevant goods. That is, for every $\valu_i$ and a relevant good $\good$, we have $
\forall \alloc_i \subseteq \goods \setminus \{\good\}, \valu_i (\alloc_i \cup \{\good\}) - \valu_i(\alloc_i)>0. 
$
For a set $S$ of goods, we define $\relevant{S}{i}$ as a subset of $S$ that are relevant to agent $i$, i.e., 
$\relevant{S}{i} = \{\good \mid \good \in S \mbox{ and } \valu_i(\{\good\})>0\}.
$
Accordingly, we define $\relevant{S}{i,j}$ as the set of goods in $S$ that are relevant both to agent $i$ and agent $j$.
These definitions are applicable to both $(p,q)$-bounded and restricted additive valuations.


\textbf{An allocation} is a partition of goods into $n+1$ bundles $\langle \alloc_{0},\alloc_1,\alloc_2,\ldots,\alloc_n \rangle$, where for every $1 \leq i \leq n$ bundle $\alloc_i$ is the share of agent $i$, and bundle $\alloc_{0}$ is the set of unallocated goods. We sometimes refer to $X_0$ as the pool. Allocation $\alloc$ is complete, if $\alloc_{0} = \emptyset$, and partial otherwise. 

Let $\beta \in [1,+\infty]$ be a constant. For a given allocation $\alloc$, we say that agent $i$ $\beta$-envies bundle $\alloc_j$ if $\beta\valu_i(\alloc_i) < \valu_i(\alloc_j)$. Additionally, agent $i$ $\beta$-strongly envies bundle $\alloc_j$ if, there exists a good $\good \in \alloc_j$, such that $\beta\valu_i(\alloc_i) < \valu_i(\alloc_j \setminus {\good})$. An allocation is $(1/\beta)$-$\efx$ if no agent $\beta$-strongly envies any other agent. For simplicity, when $\beta=1$, we drop the prefix and use terms envy, strongly envy, and $\efx$ instead of $1$-envy, $1$-strongly envy, and $1$-$\efx$.
We also consider another relaxation of the $\efx$  known as $\eftx$. In an $\eftx$ allocation, any envy that an agent $i$ may have towards $\alloc_j$ must disappear after removing any two goods from $\alloc_j$, for every $i,j$, we have
$
\forall_{\good, \good' \in \alloc_j} ,\valu_i(\alloc_i) \geq \valu_i(\alloc_j \setminus \{\good,\good'\}). 
$

\paragraph{\textbf{Weighted Envy Graph, Rank, and Virtual Value.}}
The weighted envy graph is a flexible structure that includes several meaningful graphs relating to $\efx$ allocations as a subgraph. 



\begin{definition} \label{weighted-envy-def}
	Given an allocation $\alloc$, the weighted envy graph $G_{\alloc}$ is a complete weighted bidirectional graph with $n$ vertices. For each pair of vertices $i$ and $j$, there is a directed edge from $i$ to $j$ with weight $w_{\scriptscriptstyle \alloc}(i,j) = \valu_i(\alloc_j) / \valu_{i}(\alloc_i)$. If $\valu_i(\alloc_j) = \valu_{i}(\alloc_i) = 0$, we ignore that edge.
\end{definition}

By definition, when $w_{\scriptscriptstyle \alloc}(i,j)>1$, agent $i$ envies agent $j$.  Note that the edges in $G_{\alloc}$ might have weight $+\infty$, which occurs when an agent values her bundle as 0. To prevent such situations, we initiate our algorithms with a special allocation, which we refer to as the basic feasible allocation.

\begin{definition}
$\alloc$ is a basic feasible allocation, if for every $i$, $|\alloc_i|=1$, and $\Pi_i \valu_i (\alloc_i)$ is maximized.
\end{definition}

Indeed, in the basic feasible allocation, we allocate one item to each agent, such that the Nash welfare is maximized. Lemma \ref{lem:bfa} states that we can assume without loss of generality that in the basic feasible allocation, every agent has a non-zero value for her allocated item.

\begin{lemmarep}\label{lem:bfa}
	Assume $\instance = (\agents,\goods,\values)$ is an allocation instance, and let $\alloc$ be a basic feasible allocation. If for some agent $i$, $\valu_{i}(\alloc_i) = 0$, then there exists another instance $\instance' = (\agents',\goods',\values')$ with $|\agents'|<|\agents|$, such that any $\alpha$-$\efx$ or $\alpha$-$\eftx$ guarantee  for $\instance'$ implies the same guarantee for $\instance$. 
\end{lemmarep}
\begin{proof}
	If for some agent $i$, $\valu_i(\alloc_i)=0$, then by Hall's theorem, 
	 we know that there exists a constrained set $\agents_c $ of agents such that $|\goods[\agents_c]|<|\agents_c|$. Since agents in $\agents_c$ value every good in $\goods \setminus \goods[\agents_c]$ zero, we can match the goods in $\goods[\agents_c]$ to $\agents_c$ (at most one good to each agent) and recursively find an allocation with the  desired guarantee for instance $\instance' = (\agents',\goods',\values')$ where 
	$\agents' = \agents \setminus \agents_c$, $\goods' = \goods \setminus \goods[\agents_c]$ and $\values'$ is a subset of $\values$ containing the valuation functions of the agents in $\agents'$. One can easily check that combining this allocation with the initial matching preserves the guarantee.  
\end{proof}

 We also define the \textbf{rank} of an agent for an allocation $\alloc$ as follows.
\begin{definition}\label{def:rank}
Given an allocation \(\alloc\), assume \(\valu_i(\alloc_i) > 0\) for every agent \(i\), and that \(G_{\alloc}\) has no cycle with a total product of edge weights greater than 1. The rank of agent \(i\) in \(G_{\alloc}\), denoted by \(\rank_i(\alloc)\), is defined as the maximum product of edge weights along any path ending at vertex \(i\), i.e., 
$\rank_i(\alloc) = \max_{p \in P_i} \prod_{e \in p} w_{\scriptscriptstyle \alloc}(e),
$
where \(P_i\) is the set of all paths in \(G_{\alloc}\) ending at vertex \(i\). 
Also, we define the \textbf{rankpath} of agent \(i\) as 
$\arg\max_{p \in P_i} \prod_{e \in p} w_{\scriptscriptstyle \alloc}(e),
$
and the first vertex in the rankpath of vertex \(i\) is referred to as the root of vertex \(i\), denoted by $\roott{i}{\alloc}$. If there are multiple such paths, we select the path with the fewest number of edges, and if there are still multiple  options, we select the one whose root has the smallest index. By definition, for every $i$, $\roott{i}{\alloc}$ has rank $1$.
\end{definition}

 The idea of rank is very helpful in our allocation algorithms. This concept was initially introduced by Farhadi \etal \cite{farhadi2021almost}. Here, we explore this concept and uncover several intriguing implications of rank. In particular, we employ this concept to define the virtual value of each agent.

\begin{definition} \label{virtual-value-def}
		Let
		$\alloc$ be an allocation such that for every agent $i$, $\valu_i(\alloc_i)>0$ and also $G_{\alloc}$ has no cycle with a total product of edges greater than $1$. 
		We define the virtual value of $i$ for allocation $\alloc$ as 
		$
		\virtualvalue_i(\alloc) = {\valu_i(\alloc_i)}/{\rank_i(\alloc)}
		.$
\end{definition}

Note that the virtual value of each agent is also dependent on the bundle of other agents. 
Our algorithms are designed in a way that each agent compares their virtual value with the actual values of the other agents. Accordingly, we say agent $i$ \textbf{virtually envies} agent $j$ in allocation $\alloc$, if $\virtualvalue_i(\alloc)<\valu_i(\alloc_j)$. In the same way, we define terms \textbf{virtually strong envy} and \textbf{virtually $\efx$}.


Definitions \ref{def:rank} and \ref{virtual-value-def} are well-defined only for allocations where the weighted envy graph has no cycles with a total product of weights greater than 1. Thus, it is essential to ensure that this property holds throughout our algorithms. Here, we provide a necessary and sufficient condition for $G_\alloc$ to satisfy this property. Consider the following linear program:
\begin{align} 
	\text{Minimize} \qquad&\sum_{i=1}^{n} \rupp_i \nonumber\\
	\forall_{1 \leq i,j \leq n} \qquad &\rupp_i w_{\scriptscriptstyle \alloc}(i,j) \leq \rupp_j \nonumber \\
	\forall_{1 \leq i \leq n}\qquad &\rupp_i \geq 1  \label{lp1}
\end{align}
where $w_{\scriptscriptstyle \alloc}(i,j)$ is the weight of the edge from $i$ to $j$ in $G_\alloc$. We prove that if LP\eqref{lp1} is feasible, then $G_\alloc$ has no cycle with a weight product strictly greater than 1.

\begin{lemmarep} \label{upper-bound-no-cycle}
	Given an allocation $\alloc$ such that LP\eqref{lp1} is feasible for $G_\alloc$. Then, $G_\alloc$ admits no cycle with a total weight product greater than 1.
\end{lemmarep}
\begin{proof}
	Consider a cycle $i_1 \rightarrow i_2 \rightarrow \dots \rightarrow i_k \rightarrow i_1$ in $G_\alloc$. For each edge $i_l \rightarrow i_{l + 1}$, we have $w_{\scriptscriptstyle \alloc}(i_l, i_{l + 1}) \le \rupp_{i_{l + 1}} / \rupp_{i_l}$. Therefore:
	$\prod_{l=1}^{k} w_{\scriptscriptstyle \alloc}(i_l, i_{l + 1}) \le \prod_{l=1}^{k} \frac{\rupp_{i_{l + 1}}}{\rupp_{i_l}} = 1$.
	Thus, the product of weights along any cycle is at most 1.
\end{proof}

\begin{lemmarep} \label{rank-upper-bound}
	Given an allocation $\alloc$ and let $ \rupp_1,\rupp_2,\ldots,\rupp_n$ be a feasible solution to LP\eqref{lp1}. Then, for every agent $i$ we have $\rank_i(\alloc) \leq \rupp_i$.
\end{lemmarep}
\begin{proof} 
	Consider $G_\alloc$ and let $i_1 \rightarrow i_2 \rightarrow \dots \rightarrow i_k$ be the rankpath of agent $i$. For each edge $i_l \rightarrow i_{l+1}$, we have $\rupp_{i_l}  w_{\scriptscriptstyle \alloc}(i_l, i_{l+1}) \leq \rupp_{i_{l+1}}$. Furthermore since this path is the rankpath of agent $i$ we have $\rank_{i_l}(\alloc) w_{\scriptscriptstyle \alloc}(i_l, i_{l+1}) = \rank_{i_{l+1}}(\alloc)$ and $\rank_{i_1}(\alloc) = 1$. Therefore we have:
	\begin{align*}
		\rank_i(\alloc) &= \prod_{l=1}^{k-1} w_{\scriptscriptstyle \alloc}(i_l, i_{l + 1}) & \mbox{$w_{\scriptscriptstyle \alloc}(i_l, i_{l+1}) = \rank_{i_{l+1}}(\alloc) / \rank_{i_l}(\alloc)$} \\
		&\le \prod_{l=1}^{k-1} \rupp_{i_{l+1}} / \rupp_{i_l} & \mbox{$w_{\scriptscriptstyle \alloc}(i_l, i_{l+1}) \le \rupp_{i_{l+1}} / \rupp_{i_l}$} \\
		&= \rupp_{i_k} / \rupp_{i_1} \le \rupp_i.
	\end{align*}
\end{proof}

\begin{observation} \label{obs:rankpath}
	Let $P$ be the rankpath of agent $i$. Then for every edge $j \rightarrow k \in P$ we have $\rank_j(\alloc) w_\alloc(j, k) = \rank_k(\alloc)$.
\end{observation}

Indeed, Lemmas \ref{rank-upper-bound} and \ref{upper-bound-no-cycle} together establish that if LP\eqref{lp1} is feasible for allocation $\alloc$, then  \( G_\alloc \) contains no cycle with a weight product exceeding one. Moreover, the solution to this LP corresponds precisely to the ranks of the agents. Specifically, for the restricted additive valuations, a straightforward condition on the allocation ensures the feasibility of LP \eqref{lp1}.

\begin{lemmarep}\label{lem:11}
	Suppose that the set of valuations is restricted additive, and let $\alloc$ be an allocation.
	If for each agent \( i \), \( \relevant{\alloc_i}{i} = \alloc_i \), 	the product of weights along any cycle in $G_\alloc$ is at most 1.
\end{lemmarep}
\begin{proof} 
	Let $i_1\rightarrow i_2 \rightarrow \ldots \rightarrow i_k \rightarrow i_{k+1} = i_1$ be a cycle in $G_\alloc$. We have 
	\begin{align*}
		\prod_{l=1}^{k} w_{\scriptscriptstyle \alloc}(i_l,i_{l+1}) &= \frac{\valu_{i_1}(\alloc_{i_2})}{\valu_{i_1}(\alloc_{i_1})}\cdot \frac{\valu_{i_2}(\alloc_{i_3})}{\valu_{i_2}(\alloc_{i_2})}\cdot \ldots \cdot \frac{\valu_{i_k}(\alloc_{i_1})}{\valu_{i_k}(\alloc_{i_k})}\\
		&\leq\frac{\valu_{i_1}(\alloc_{i_2})}{\valu_{i_2}(\alloc_{i_2})}\cdot \frac{\valu_{i_2}(\alloc_{i_3})}{\valu_{i_3}(\alloc_{i_3})}\cdot \ldots \cdot \frac{\valu_{i_k}(\alloc_{i_1})}{\valu_{i_1}(\alloc_{i_1})}\\
		&=\frac{\valu_{i_1}(\alloc_{i_2})}{\valu(\alloc_{i_2})}\cdot \frac{\valu_{i_2}(\alloc_{i_3})}{\valu(\alloc_{i_3})}\cdot \ldots \cdot \frac{\valu_{i_k}(\alloc_{i_1})}{\valu(\alloc_{i_1})}\\
		&\leq 1.
	\end{align*}
\end{proof}


In this paper, we use $G_{\beta, \alloc}$ to refer to a specific subgraph of $G_{\alloc}$ that comprises all the vertices but includes only the edges with weight more than $\beta$. In Sections \ref{sec:7}, \ref{sec:5}, and \ref{sec:8} we respectively use $G_{1,\alloc}$, $G_{\sqrt 2,\alloc}$, and $G_{0,\alloc}$ as our base graphs. In Lemma \ref{lem:12}
 we prove a structural property of $G_{\beta,\alloc}$.

\begin{lemmarep}\label{lem:12}
	Consider restricted additive valuations, and let $\alloc$ be a $1/\beta$-$\efx$ allocation for some $\beta \in [1,+\infty)$. For every edge $i \rightarrow j$ in $G_{\beta,\alloc}$, it follows that $\valu_i(\alloc_j) = \valu(\alloc_j)$.
\end{lemmarep}
\begin{proof}
	Edge $i\rightarrow j$ in $G_{\beta,\alloc}$ implies $\beta\valu_i(\alloc_i)<\valu_{i}(\alloc_j)$. Therefore, if there's a good $\good$ with value $0$ to agent $i$ in $\alloc_j$, we have $\beta\valu_i(\alloc_i)<\valu_{i}(\alloc_j \setminus\{\good\})$, which is in contradiction with the $1/\beta$-$\efx$ property of $\alloc$.
\end{proof}

\begin{figure}
\centering
\begin{minipage}{5cm}
\tikzset{every picture/.style={line width=0.75pt}} 

\tikzset{every picture/.style={line width=0.75pt}} 

\tikzset{every picture/.style={line width=0.75pt}} 

\tikzset{every picture/.style={line width=0.75pt}} 

\tikzset{every picture/.style={line width=0.75pt}} 

\begin{tikzpicture}[x=0.75pt,y=0.75pt,yscale=-1,xscale=1]
	\draw [color={rgb, 255:red, 0; green, 0; blue, 0 }  ,draw opacity=1 ][line width=0.75]    (290,70) -- (361,70) ;
	\draw [color={rgb, 255:red, 0; green, 0; blue, 0 }  ,draw opacity=1 ][line width=0.75]    (361,141) -- (361,70) ;
	\draw [color={rgb, 255:red, 0; green, 0; blue, 0 }  ,draw opacity=1 ][line width=0.75]    (361,141) -- (290,70) ;
	\draw [color={rgb, 255:red, 0; green, 0; blue, 0 }  ,draw opacity=1 ][line width=0.75]    (361,141) -- (290,141) ;
	\draw [color={rgb, 255:red, 0; green, 0; blue, 0 }  ,draw opacity=1 ][line width=0.75]    (290,141) -- (361,70) ;
	\draw [color={rgb, 255:red, 0; green, 0; blue, 0 }  ,draw opacity=1 ][line width=0.75]    (290,141) -- (290,70) ;
	\draw [color={rgb, 255:red, 0; green, 0; blue, 0 }  ,draw opacity=1 ][line width=0.75]    (290,70) .. controls (269,70.33) and (270,140.33) .. (290,141) ;
	\draw  [color={rgb, 255:red, 0; green, 0; blue, 0 }  ,draw opacity=1 ][fill={rgb, 255:red, 0; green, 0; blue, 0 }  ,fill opacity=1 ][line width=0.75]  (286,70) .. controls (286,67.79) and (287.79,66) .. (290,66) .. controls (292.21,66) and (294,67.79) .. (294,70) .. controls (294,72.21) and (292.21,74) .. (290,74) .. controls (287.79,74) and (286,72.21) .. (286,70) -- cycle ;
	\draw  [color={rgb, 255:red, 0; green, 0; blue, 0 }  ,draw opacity=1 ][fill={rgb, 255:red, 0; green, 0; blue, 0 }  ,fill opacity=1 ][line width=0.75]  (357,70) .. controls (357,67.79) and (358.79,66) .. (361,66) .. controls (363.21,66) and (365,67.79) .. (365,70) .. controls (365,72.21) and (363.21,74) .. (361,74) .. controls (358.79,74) and (357,72.21) .. (357,70) -- cycle ;
	\draw  [color={rgb, 255:red, 0; green, 0; blue, 0 }  ,draw opacity=1 ][fill={rgb, 255:red, 0; green, 0; blue, 0 }  ,fill opacity=1 ][line width=0.75]  (286,141) .. controls (286,138.79) and (287.79,137) .. (290,137) .. controls (292.21,137) and (294,138.79) .. (294,141) .. controls (294,143.21) and (292.21,145) .. (290,145) .. controls (287.79,145) and (286,143.21) .. (286,141) -- cycle ;
	\draw  [color={rgb, 255:red, 0; green, 0; blue, 0 }  ,draw opacity=1 ][fill={rgb, 255:red, 0; green, 0; blue, 0 }  ,fill opacity=1 ][line width=0.75]  (357,141) .. controls (357,138.79) and (358.79,137) .. (361,137) .. controls (363.21,137) and (365,138.79) .. (365,141) .. controls (365,143.21) and (363.21,145) .. (361,145) .. controls (358.79,145) and (357,143.21) .. (357,141) -- cycle ;
	\draw (284.8,51.2) node [anchor=north west][inner sep=0.75pt]  [font=\scriptsize]  {$1$};
	\draw (354.8,51.2) node [anchor=north west][inner sep=0.75pt]  [font=\scriptsize]  {$2$};
	\draw (355.8,144.2) node [anchor=north west][inner sep=0.75pt]  [font=\scriptsize]  {$4$};
	\draw (284.8,144.2) node [anchor=north west][inner sep=0.75pt]  [font=\scriptsize]  {$3$};
	\draw (362.2,91.4) node [anchor=north west][inner sep=0.75pt]  [font=\scriptsize]  {$7$};
	\draw (327.6,80) node [anchor=north west][inner sep=0.75pt]  [font=\scriptsize]  {$6$};
	\draw (317.6,55.2) node [anchor=north west][inner sep=0.75pt]  [font=\scriptsize]  {$10$};
	\draw (290.4,91.4) node [anchor=north west][inner sep=0.75pt]  [font=\scriptsize]  {$3$};
	\draw (261,91.4) node [anchor=north west][inner sep=0.75pt]  [font=\scriptsize]  {$3$};
	\draw (314.6,139.2) node [anchor=north west][inner sep=0.75pt]  [font=\scriptsize]  {$10$};
	\draw (328.6,116) node [anchor=north west][inner sep=0.75pt]  [font=\scriptsize]  {$6$};
\end{tikzpicture}
\end{minipage}
\begin{minipage}{5cm}
\begin{tabular}{|l|l|l|l|l|l|l|l|}
	\hline
	& $\good_1$ & $\good_2$ & $\good_3$ & $\good_4$ & $\good_5$ & $\good_6$  & $\good_7$  \\ \hline
	$v_1$ & 3 & 3 & 6 & 0 & 0 & 10 & 0  \\ \hline
	$v_2$ & 0 & 0 & 0 & 6 & 7 & 10 & 0  \\ \hline
	$v_3$ & 3 & 3 & 0 & 6 & 0 & 0  & 10 \\ \hline
	$v_4$ & 0 & 0 & 6 & 0 & 7 & 0  & 10 \\ \hline
\end{tabular}
\end{minipage}
\caption{An example of valuations that are both restricted additive and $(2, 2)$-bounded}
\label{fig:tb1}
\end{figure}
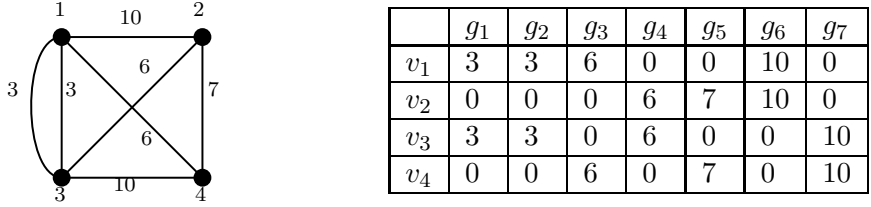
\begin{example}\label{ex-1}
	Consider an instance with $4$ agents and $7$ goods. The (additive) values of the agents for these goods are shown in Figure \ref{fig:tb1}. This set of valuation functions is restricted additive with the following inherent values: $\valu_{\good_1} =  \valu_{\good_2} = 3$, $\valu_{\good_3} = \valu_{\good_4} = 6$, $\valu_{\good_5} = 7$, and $\valu_{\good_6} = \valu_{\good_7} = 10$. In addition, this set of valuation functions is $(2,2)$-bounded, meaning that each good is relevant to at most two agents, and each pair of agents has at most two relevant goods. 
	Indeed, agents 1 and 3 are the only agents with two common relevant goods, and every other pair of agents has at most one common relevant good. 	So if we remove $\good_1$, the instance becomes $(2, 1)$-bounded.
	Now, consider the allocation depicted in Figure \ref{fig:33}. In this allocation, we have 
	$\alloc_{0} = \{ \good_5 \}$, 
	$\alloc_{1} = \{ \good_1, \good_3 \}$, 
	$\alloc_{2} = \{ \good_6 \}$, 
	$\alloc_{3} = \{ \good_2, \good_4 \}$, 
	$\alloc_{4} = \{ \good_7 \}$, 
	and values are 
	$\valu({\alloc_1}) = 9$, 
	$\valu({\alloc_2}) = 10$, 
	$\valu({\alloc_3}) = 9$, 
	$\valu({\alloc_4}) = 10$.  
	Here, for every agent $i$ we have $\valu_i(\alloc_i) = \valu({\alloc_i})$ which means that no good with value $0$ is allocated to any agent. In Figure \ref{fig:33} you can  find graphs $G_{\alloc}$,  $G_{0,\alloc}$, $G_{1,\alloc}$ and $G_{0.6,\alloc}$. As you can see $G_{0,\alloc}$, $G_{0.5,\alloc}$ and $G_{1,\alloc}$ are subgraph of $G_\alloc$ containing only edges with value respectively more than $0$, $0.5$ and $1$. 
	Now, consider $G_{0,\alloc}$. This graph contains multiple paths that lead to vertex $4$, such as $p_1 = 3 \rightarrow 4$ and $p_2 = 1 \rightarrow 2 \rightarrow 3 \rightarrow 4$. The product of the edge weights in path $p_1$ is $10/9$, while the product of the edge weights in path $p_2$ is $20/27$. By comparing the products of weights of different paths leading to vertex $4$, we can see that path $p_1$ has the highest product of weights. Therefore, $\rank_4(\malloc) = 10/9$ and $\virtualvalue_4(\malloc) = 9$. 
	
	\input{src/fig_alloc}	
	\begin{figure}
\centering       

\begin{minipage}{5cm}

\tikzset{every picture/.style={line width=0.45pt}} 

\begin{tikzpicture}[x=0.75pt,y=0.75pt,yscale=-1,xscale=1]
	
	\draw  [fill={rgb, 255:red, 0; green, 0; blue, 0 }  ,fill opacity=1 ] (92,65.5) .. controls (92,63.57) and (93.57,62) .. (95.5,62) .. controls (97.43,62) and (99,63.57) .. (99,65.5) .. controls (99,67.43) and (97.43,69) .. (95.5,69) .. controls (93.57,69) and (92,67.43) .. (92,65.5) -- cycle ;
	\draw  [fill={rgb, 255:red, 74; green, 144; blue, 226 }  ,fill opacity=0.3 ] (83.75,119.5) -- (164,119.5) -- (164,132) -- (83.75,132) -- cycle ;
	\draw  [fill={rgb, 255:red, 208; green, 2; blue, 27 }  ,fill opacity=0.32 ] (147.75,55.25) -- (161,55.25) -- (161,74.5) -- (147.75,74.5) -- cycle ;
	\draw  [fill={rgb, 255:red, 184; green, 233; blue, 134 }  ,fill opacity=0.3 ] (165.03,58.97) -- (164.93,72.04) -- (102.14,71.58) -- (101.68,134.56) -- (87.79,134.46) -- (88.35,58.41) -- cycle ;
	\draw  [fill={rgb, 255:red, 0; green, 0; blue, 0 }  ,fill opacity=1 ] (151.25,65.25) .. controls (151.25,63.32) and (152.82,61.75) .. (154.75,61.75) .. controls (156.68,61.75) and (158.25,63.32) .. (158.25,65.25) .. controls (158.25,67.18) and (156.68,68.75) .. (154.75,68.75) .. controls (152.82,68.75) and (151.25,67.18) .. (151.25,65.25) -- cycle ;
	\draw  [fill={rgb, 255:red, 0; green, 0; blue, 0 }  ,fill opacity=1 ] (91.75,125.25) .. controls (91.75,123.32) and (93.32,121.75) .. (95.25,121.75) .. controls (97.18,121.75) and (98.75,123.32) .. (98.75,125.25) .. controls (98.75,127.18) and (97.18,128.75) .. (95.25,128.75) .. controls (93.32,128.75) and (91.75,127.18) .. (91.75,125.25) -- cycle ;
	\draw  [fill={rgb, 255:red, 0; green, 0; blue, 0 }  ,fill opacity=1 ] (151.75,125.5) .. controls (151.75,123.57) and (153.32,122) .. (155.25,122) .. controls (157.18,122) and (158.75,123.57) .. (158.75,125.5) .. controls (158.75,127.43) and (157.18,129) .. (155.25,129) .. controls (153.32,129) and (151.75,127.43) .. (151.75,125.5) -- cycle ;
	
	\draw (74.5,52.4) node [anchor=north west][inner sep=0.75pt]  [font=\footnotesize]  {$1$};
	\draw (167,53.4) node [anchor=north west][inner sep=0.75pt]  [font=\footnotesize]  {$2$};
	\draw (75,134.9) node [anchor=north west][inner sep=0.75pt]  [font=\footnotesize]  {$3$};
	\draw (162.75,135.15) node [anchor=north west][inner sep=0.75pt]  [font=\footnotesize]  {$4$};
	\draw (73.25,87.4) node [anchor=north west][inner sep=0.75pt]  [font=\scriptsize]  {$g_{1}$};
	\draw (120,135.65) node [anchor=north west][inner sep=0.75pt]  [font=\scriptsize]  {$g_{2}$};
	\draw (149.75,77.9) node [anchor=north west][inner sep=0.75pt]  [font=\scriptsize]  {$g_{3}$};
\end{tikzpicture}
\end{minipage}
\begin{minipage}{5cm}
	\begin{tabular}{|l|l|l|l|}
		\hline
		& $g_1$ & $g_2$ & $g_3$ \\ \hline
		$v_1$ & 2     & 0     & 0     \\ \hline
		$v_2$ & 1     & 0     & 5     \\ \hline
		$v_3$ & 4     & 3     & 0     \\ \hline
		$v_4$ & 0     & 4     & 0     \\ \hline
	\end{tabular}
\end{minipage}
\caption{Example of $(\infty,1)$-bounded valuations}
\label{fig:4}
\end{figure}
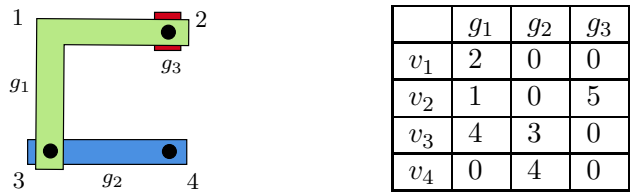
	
\end{example}
\begin{example}
	In the instance depicted in Figure \ref{fig:4}, there are four agents with additive valuations and three goods. The value of good $\good_1$ varies across agents, indicating that the valuations are not restricted additive. Since good $\good_1$ is relevant to three agents, we have $p=3$. Moreover,  for each pair of goods, there is at most one agent that is relevant to both of them. Therefore, the set of valuation functions is $(3,1)$-bounded. The hypergraph corresponding to this set of valuation functions is depicted in Figure \ref{fig:4}.
\end{example}

\begin{remark}
	The allocation shown in Example \ref{ex-1}  can serve as an evidence illustrating that the $\efx$ with charity allocation algorithm proposed by Akrami et al. \cite{akrami2022ef2x} might terminate before allocating the whole set of goods. To establish the algorithm's termination, they introduce a potential function $\phi(\cdot)$ and demonstrate that, with each update, the potential function increases. This particular allocation achieves the maximum value of $\phi$ among all possible allocations. Consequently, any modification to this allocation decreases the value of $\phi$. However, the allocation is not complete.
\end{remark}

\subsection{ A General Framework of Our Allocation Algorithms}
Although we present algorithms with different guarantees for various valuation functions, they all follow a unified framework, which we describe here to avoid repetition.

Each algorithm consists of a set of updating rules, referred to by their indices (e.g., Rule 1, Rule 2, etc.). Every rule has preconditions that determine when it applies and updates that modify the allocation. The process starts with a basic feasible allocation, and at each step, the lowest applicable rule is selected and applied.

We also define several properties and ensure that these properties are maintained after each update rule is applied. Additionally, we show that these properties hold for basic feasible allocations.
For each algorithm, we define a potential function $\phi$ and show that its value increases after each update rule is applied. With a known upper bound on the value of this function, we are guaranteed that the algorithm will eventually reach a state where no more updates can be made. At this point, we perform a final step and return the resulting allocation. We then show that the final allocation satisfies the desired fairness notion.

Algorithm \ref{frame} shows the pseudocode of our general framework. We use $\malloc$ and $\malloc'$ to refer to our allocation before and after each update, respectively. Also, we denote the final allocation with $\malloc^*$.

\begin{algorithm}[t]
	\caption{General Framework for Allocation Algorithms}
	\label{frame}
	\SetAlgoLined
	\SetKwFunction{FMain}{Allocate}
	\SetKwProg{Fn}{Function}{:}{}
	\Fn{\FMain{$\malloc_0$}}{
		 $\malloc \gets \mbox{\small \textsf{Basic Feasible Allocation}}$\;
		\While{there exists an applicable update rule }{
			\tcp{The number of rules is $k$}
			\For{$i \gets 1$ \KwTo $k$ }{
				\If{Rule $i$ is applicable on $\malloc$}{
					Apply Rule $i$ to $\malloc$ to obtain $\malloc'$\;
					$\malloc \gets \malloc'$\;
					\textbf{break};
				}
			}
		}
		Execute final step on $\malloc$ to obtain $\malloc^*$\;
		\Return $\malloc^*$;
	}
\end{algorithm}

\section{Results Overview}\label{sec:resultstechniques}
We provide fairness guarantees for restricted additive and $(p,q)$-bounded valuations. These results, alongside existing results for the restricted additive and $(p,q)$-bounded valuations form a symmetrical pattern:
\begin{itemize} 
	\item Akrami \etal \cite{akrami2022ef2x} show that for $(\infty, \infty)$-bounded restricted additive valuations a complete $\eftx$ allocation, and a partial $\efx$ allocation with less than $n/2$ discarded goods exists. We prove the same guarantees for $(\infty, 1)$-bounded valuations (Theorems \ref{thm7} and \ref{thm7f}). 
	
	\item Christodoulou \cite{christodoulou2023fair} show that for $ (2,1)$-bounded valuations we can find an $\efx$ allocation. We prove the same guarantee for $ (2, \infty)$-bounded restricted additive valuations (Theorem \ref{thm8}). 
	\item We show that both $ (\infty, 1)$-bounded subadditive valuations, and $(\infty,\infty)$-bounded valuations with the restricted additive property admit a $({\sqrt 2}/{2})$-$\efx$ allocation (Theorems \ref{thm:fstep} and \ref{thm72}). 
\end{itemize}

\begin{table}[t]
	\onehalfspacing
	\centering
	\begin{tabular}{|c|cc|}
		\hline
		& \multicolumn{1}{c|}{Monotone} & Restricted additive \\ \hline
		(2,1)-bounded        & \multicolumn{2}{c|}{$\efx$ \cite{christodoulou2023fair} \textcolor{cyan}{$\vardiamondsuit$}}\\ \hline
		(2,$\infty$)-bounded & \multicolumn{1}{c|}{Open}        & $\efx$ (Theorem \ref{thm8})\textcolor{magenta}{$\vardiamondsuit
			$}              \\ \hline
		($\infty$,1)-bounded &
		\multicolumn{2}{c|}{\begin{tabular}[c]{@{}c@{}}$\eftx$ (Theorem \ref{thm7f})\textcolor{cyan}{$\clubsuit
					$}\\ $\langle \efx,\lfloor \frac{n}{2}\rfloor -1 \rangle$ (Theorem \ref{thm7})\textcolor{cyan}{$\spadesuit
					$} \\ 
				$\frac{\sqrt{2}}{2}$-$\efx$ (Theorem \ref{thm72}$^*$)\textcolor{cyan}{$\varheartsuit
					$}\end{tabular}} \\ \hline
		\multicolumn{1}{|l|}{($\infty$,$\infty$)-bounded} &
		\multicolumn{1}{c|}{Open} &
		\begin{tabular}[c]{@{}c@{}}$\eftx$ ~\cite{akrami2022ef2x}\textcolor{magenta}{$\clubsuit
				$}\\ $\langle \efx,\lfloor \frac{n}{2}\rfloor -1 \rangle$ ~\cite{akrami2022ef2x}\textcolor{magenta}{$\spadesuit
				$}\\ $\frac{\sqrt{2}}{2}$-$\efx$ (Theorem \ref{thm:fstep})\textcolor{magenta}{$\varheartsuit
				$}\end{tabular} \\ \hline
	\end{tabular}
	\vspace{0.2cm}
	\caption{
		Overview of the results. To show the symmetry of the results for two valuations, we use symbols $\clubsuit, \vardiamondsuit, \spadesuit, \varheartsuit$. Results with the same label are symmetric. 
		Note that for Theorem \ref{thm72}, we also have a subadditivity assumption.
	}
	\label{tab:my-table}
\end{table}
In our algorithms across different sections, we use various subgraphs of the weighted envy graph, including $G_{0,\alloc}$, $G_{1,\alloc}$, and $G_{\sqrt{2},\alloc}$. For all these algorithms, it is crucial to ensure that $G_\alloc$ does not contain a cycle with a total product of weights strictly greater than 1.
For the restricted additive valuations, a simple property guarantees this condition. As shown in Lemma \ref{lem:11}, if we allocate only relevant goods to each agent, then $G_\alloc$ does not have such cycles (Properties \ref{prop:2} and \ref{prop42}).  For $(p,q)$-bounded valuations, we perform our update rules in a way that keeps LP\eqref{lp1} feasible (Property \ref{q1-no-greater-1-cycle}). 
When $G_\alloc$ has no cycle with a total product of weights strictly greater than one, terms such as rank, rankpath, and virtual value are well-defined. Another property that we maintain in our algorithms is that the \textbf{allocation is $\efx$ with respect to the virtual values} (Properties \ref{prop43} and \ref{q1-virtually-efx} ). Note that this property is stronger than the $\efx$ property. 


As a warmup, in Section \ref{sec:7} we show that every family of $(\infty,1)$-bounded valuation functions admit $\efx$ allocations with less than $n/2$ discarded goods. Next, we extend this allocation and obtain a complete $\eftx$ allocation. These two results complement the guarantees of Akrami \etal \cite{akrami2022ef2x} for the restricted additive setting.  Hence, most of the rules we use in our allocation algorithms resemble those in Akrami \etal \cite{akrami2022ef2x}. However, the structure of $(p,q)$-bounded valuations allows us to use a much simpler potential function to prove the termination of our algorithms. 

An essential part of the algorithm proposed by Akrami \etal \cite{akrami2022ef2x} is an elegant process called envy-elimination. In the envy-elimination process, we eliminate all possible strong envies from the allocation while ensuring that the potential function either remains unchanged or increases. Surprisingly, in the envy-elimination process we only remove goods from the bundles, however, by carefully defining the potential function, we can show that we never fall into an infinite loop trap. 
The reason that here our potential function is much simpler is that, here we don't use envy-elimination in our algorithm. Instead, we make sure that after each update the allocation is $\efx$.

The envy-elimination process makes the restricted additive approach a great starting point for deriving more general $\efx$ results: for restricted additive valuations, we needn't worry about the $\efx$ guarantee after each update, as envy-elimination restores this property. This allows for various updates. Here, we refine the envy-elimination process to obtain a $(\sqrt{2}/2)$-$\efx$ allocation for restricted additive valuations. Our algorithm uses two simple update rules, envy-elimination, and a final step.

Interestingly, we achieve the same guarantee for $(\infty,1)$-bounded subadditive valuations using a simple algorithm consisting of just \textbf{one updating rule} and a final step. This rule heavily relies on concepts such as rank and virtual value. In this update rule, if any agent virtually envies the pool, we select a subset of the goods in the pool and choose an appropriate agent based on the ratio of the value of that subset to the agent's virtual value.  We then allocate this subset to the selected agent and adjust the allocation through her rankpath. This update rule turns out to be very useful.

We later leverage the same rule to propose an $\efx$ allocation algorithm for the case that $p=2$ and the valuations are restricted additive. Our algorithm and analysis for this setting is more technically involved and the algorithm includes four updating rules and a final step. However, the key updating rule in this algorithm remains similar to the previous algorithm.
Since our last two algorithms do not involve envy-elimination, their potential functions are simple:  after each update, the Nash welfare increases.

 We note that almost all previously known updating rules for $\efx$ allocations for a general number of agents are designed to increase the utility of agents. However, for special cases, such as when $n=3$, some update rules might actually decrease social welfare. This discrepancy suggests that for the general case, we may need updating rules that allow for a temporary decrease in agents' utility to successfully perform the algorithm. For the restricted additive setting, Akrami \etal  \cite{akrami2022ef2x} employed the envy elimination process to reduce agents' utility. By employing a complex potential function, they show that the algorithm never falls into an infinite loop. In our approach, we leverage the concept of virtual value to propose updates that can decrease the value of agents while ensuring that the algorithm does not fall into an infinite loop. This  is why we believe that the idea of virtual value has the potential to be applied to more general cases. 

Several intriguing directions emerge for further exploration following this work. One key area of interest is whether the observed symmetry between restricted additive valuations and $(\infty,1)$-bounded valuations extends to other problem domains. Notably, there are significant results for the maximin fairness notion and restricted additive valuations, commonly referred to as the Santa Claus problem \cite{asadpour2012santa,annamalai2017combinatorial,bamas2023better}. Investigating whether analogous results can be achieved for the Santa Claus problem with $(\infty,1)$-bounded valuations would be of great interest.
Furthermore, it is worthwhile to consider the generalization of our $\frac{\sqrt{2}}{2}$ algorithms to more general settings, such as additive and subadditive valuations.  One can think of using the concept of rank and virtual value and the idea in the suggested updating rule to potentially achieve a $\frac{\sqrt{2}}{2}$-$\efx$ allocation in the additive setting.

	\section{Further Related Work}\label{sec:related}



\paragraph{Finding approximate $\efx$ allocations.} Finding approximately fair allocations is a common approach for various fairness notions, including $\efx$. Other than studies we mentioned in the introduction, Barman \etal~ \cite{barman2023parameterized} have explored the existence of approximate $\efx$ allocations. They introduce a parameter $\gamma \in (0,1]$ to bound the multiplicative range of nonzero values for goods across agents, and develop an algorithm that yields a $\frac{2\gamma}{\sqrt{5+4\gamma-1}}$-$\efx$ allocation.
Amanatidis $\etal$ \cite{amanatidis2024pushing} also investigated $(2,\infty)$-bounded valuations and proved the existence of a $2/3$-$\efx$ allocation. They also proved that this guarantee extends to instances where each good has only three distinct values for the agents.

\paragraph{Relaxing $\efx$.} Some studies explore relaxed variations of $\efx$ to achieve better guarantees. Farhadi \etal \cite{farhadi2021almost} examine removing a random good instead of the least valuable one, establishing a $0.74$ approximation for this modified notion. Caragiannis \etal \cite{caragiannis2022existence} introduce epistemic EFX, where for each agent, there exists a way to shuffle others' goods so the agent doesn't envy anyone. Another variation is envy-freeness up to one less preferred good, introduced by Barman \etal \cite{berger2022almost}, which limits the value of the eliminated good.

\paragraph{$\efx$ with Charity.} A strand of research has delved into finding partial $\efx$ allocations rather than complete ones. Naturally, we want to keep the number of discarded goods as small as possible. The pioneering work in this area was a partial EFX allocation, with the Nash welfare at least half of the optimal \cite{caragiannis2019envy}. Subsequently, Chaudhury \etal \cite{chaudhury2021little} prove the existence of a partial $\efx$ allocation that leaves at most $n-1$ goods unallocated. Berger \etal \cite{berger2022almost} reduced the number of discarded goods to $n-2$. Chaudhury \etal \cite{chaudhury2021improving} presented a general framework for achieving a partial $(1-\epsilon)$-$\efx$ allocations with a sub-linear number of unallocated goods for any $\epsilon$. The number of unallocated goods is then successively reduced to ultimately $O_\epsilon(\sqrt{n \log n})$ \cite{jahan2022rainbow,akrami2022efx,berendsohn2022fixed}.

Finally, recent surveys by Aziz \etal \cite{aziz2022algorithmic} and Amanatidis \etal \cite{amanatidis2022fair} provide valuable insights into the fair allocation of indivisible goods.

	\section{A Warmup: A Partial $\efx$ and a Complete $\eftx$ Allocation for $(\infty,1)$-bounded Valuations}\label{sec:7}

We start by proposing an algorithm that returns an $\eftx$ allocation for $(\infty,1)$-bounded valuations. The algorithm involves four updating rules, followed by a  final step. We show that if none of the rules is feasible, the allocation is $\efx$ with at most $\lfloor \frac{n}{2} \rfloor - 1$ number of discarded goods. Next, we show that after running the final step, the output is a complete $\eftx$ allocation. 
The majority of our updating rules are similar to those of Akrami \etal \cite{akrami2022ef2x} for the restricted additive valuations.
We design the rules such that the following potential function increases after each update:
\begin{align*}
	\phi(\malloc) = \bigg[\sum_{i=1}^{n} \valu_i(\malloc_i), \sum_{i=1}^{n} |\malloc_i| \bigg].
\end{align*}

Indeed, after each update, either the social welfare increases, or the social welfare remains unchanged while the number of allocated items increases. Our potential function is simpler compared to the one proposed by Akrami \etal \cite{akrami2022ef2x}, primarily due to the absence of the envy-elimination process in our algorithm.
In addition, we guarantee that after each update, the allocation remains $\efx$, and furthermore, the following property holds:

\begin{enumerate}[label=\textcolor{red}{$\ddag\!$} (\roman*)]
	\item \label{prop:51} For each \textbf{non-source} agent $i$ in $G_{1, \malloc}$, $\relevant{\malloc_i}{i} = \malloc_i$.

\end{enumerate}

One can easily check that the basic feasible allocation satisfies Property \ref{prop:51}. Note that the underlying assumption of Property \ref{prop:51} is that $G_{1,\malloc}$ is acyclic. To maintain this property, after each update, we remove any envy-cycle in $G_{1,\malloc}$ by rotating the bundles through that cycle. Formally, after each update, while there is a cycle $\cycle$ in $G_{1,\malloc}$, we reallocate the bundles in $C$ as follows: for each edge $i \rightarrow j \in \cycle$, we allocate $\relevant{\malloc_j}{i}$ to agent $i$ and return $\malloc_j \setminus  \relevant{\malloc_j}{i}$ back to the pool. Trivially, resolving the envy-cycles increases the potential function, and does not violate Property \ref{prop:51}. For convenience, we ensure that after each update, we resolve all envy-cycles.
 In the following sections we describe our updating rules and their properties.  We designate our algorithm in this section as $\algo$.

\subsection{Update Rule 1}

Rule 1 is feasible when there is an agent $i$ and a good $\good \in \malloc_0$ such that $\valu_i(\{\good\}) > \valu_i(\malloc_i)$. In such cases, we allocate $\good$ to agent $i$ and return the goods in $\malloc_i$ back to the pool. Next, we resolve all the envy-cycles.

\begin{lemma}\label{lem:721}
	Let $\malloc$ be an $\efx$ allocation satisfying Property \ref{prop:51}, and let $\malloc'$ be the allocation after applying Rule 1 on $\malloc$. Then $\malloc'$ is $\efx$ and satisfies Property \ref{prop:51}. Furthermore, we have $\phi(\malloc') > \phi(\malloc)$. 
\end{lemma}
\begin{proof}
		The update exclusively impacts the bundle of agent $i$. Given that $\valu_i(\malloc'_i) > \valu_i(\malloc_i)$, agent $i$ does not strongly envy any other agent in $\malloc'$. Furthermore, as $|\malloc'_i| = 1$, no other agent strongly envies agent $i$. Consequently, $\malloc'$ is $\efx$. 
		To prove the validity of Property \ref{prop:51}, it suffices to show that all goods in $\malloc'_i$ are relevant to agent $i$, which is indeed the case since $\malloc'_i = {\good}$ and the $\good$ is relevant to agent $i$. Therefore, Property \ref{prop:51} holds.
Also, this update increases social welfare and therefore, after the update we have $\phi(\malloc')> \phi(\malloc)$.
\end{proof}

\subsection{Update Rule 2}

The properties of $(\infty,1)$-bounded valuations are pivotal for this updating rule. This rule is feasible when an agent $i$ and a good $\good \in \malloc_0$ meet the following conditions:  agent $i$ is a source in $G_{1, \malloc}$, and $\good$ is relevant to agent $i$. If so, we allocate $\good$ to agent $i$ and return the goods in $\malloc_i$ that are irrelevant to agent $i$ back to the pool. In other words, we set $\malloc'_i = \relevant{(\malloc_i \cup \{\good\})}{i}$. Next, we resolve envy-cycles.

\begin{lemma}\label{lem:731}
		Let $\malloc$ be an $\efx$ allocation satisfying Property \ref{prop:51}. Assume that Rule 1 is not applicable on $\malloc$, and let $\malloc'$ be the allocation after applying Rule 2 on $\malloc$. Then $\malloc'$ is $\efx$ and satisfies Property \ref{prop:51}. Furthermore, we have $\phi(\malloc') > \phi(\malloc)$.
\end{lemma}
\begin{proof}
	The bundle of agent $i$ is the only bundle that has been affected, and we have that $\valu_i(\malloc'_i) > \valu_i(\malloc_i)$. Therefore, we only need to prove that no other agent $j$ envies agent $i$. 
	Note that since $i$ has no incoming edge in $G_{1,\malloc}$, agent $j$ does not envy any single good in $\malloc_i$. Furthermore,  since Rule 1 is not applicable, agent $j$ does not envy good $\good$. In addition, since all the goods in $\malloc'_i$ are relevant to agent $i$ and the valuations are $(\infty, 1)$-bounded, at most one good in $\malloc'_i$ is relevant to $j$. Hence, 
	\begin{align*}
		\valu_j(\malloc'_j) &= \valu_j(\malloc_j) &\mbox{$\malloc'_j = \malloc_j$} \\
		&\ge \max_{\good' \in \malloc'_i} \: \space \valu_j(\{\good'\}) &\mbox{$j$ does not envy any good in $\malloc'_i$} \\
		&= \valu_j(\malloc'_i) &\mbox{valuations are $(\infty, 1)$-bounded}. \\
	\end{align*}
	Therefore, the resulting allocation is $\efx$.	
	Also, since all the goods in the bundle of agent $i$ are relevant to her, Property \ref{prop:51} holds. 
Finally, this update increases social welfare and therefore, after the update we have $\phi(\malloc') > \phi(\malloc)$.
\end{proof}

\subsection{Update Rule 3}
This rule is feasible when there exist an agent $i$ and a good $\good \in \malloc_0$ such that $i$ is a source in $G_{1, \malloc}$ and there is no other agent $j$ such that $\valu_j(\malloc_j) < \valu_j(\malloc_i \cup \{\good\})$. In that case, we allocate good $\good$ to agent $i$ and resolve all the envy-cycles.

\begin{lemma}\label{lem:741}
		Let $\malloc$ be an $\efx$ allocation satisfying Property \ref{prop:51}. Assume that Rule 1 and Rule 2 are not applicable on $\malloc$, and let $\malloc'$ be the allocation after applying Rule 3 on $\malloc$. Then $\malloc'$ is $\efx$ and satisfies Property \ref{prop:51}. Furthermore, we have  $\phi(\malloc') > \phi(\malloc)$.
\end{lemma}
\begin{proof}
	According to preconditions of this update rule, after allocating good $\good$ to agent $i$, no other agent envies her, which means allocation $\malloc'$ is $\efx$.  In addition, agent $i$ remain a source and does not violate Property \ref{prop:51}.
Furthermore, since Update Rule 2 is not applicable, good $\good$ is irrelevant to agent $i$. Therefore, the social welfare of the agents remains unchanged after this allocation, while the total number of allocated goods increases. As a result, we have $\phi(\malloc') > \phi(\malloc)$.
\end{proof}

\subsection{Update Rule 4}

To describe our last updating rule, we need first to define the concept of the most envious agent.
	Suppose we have an allocation $\alloc$, let $S$ be a set of goods,
	and suppose there are one or more agents who envy $\sgoods$. We can identify the minimal subset of $\sgoods$, referred to as $\sgoods'$, which is envied by at least one agent, namely $i$. Consequently, if we allocate the set $\sgoods'$ to agent $i$, the value of her bundle increases. Furthermore, no other agent strongly envies agent $i$ since there is no agent who envies a strict subset of $\sgoods'$. We call agent $i$, the most envious agent of set $S$. Note that there might be multiple options as the most envious agent for a set $S$.

Denote the sources of $G_{1,\malloc}$ by $s_1, s_2, \dots, s_k$, and define $\comp_{\malloc,s_i}$ as the set of agents that can be reached from $s_i$ via a directed path in $G_{1,\malloc}$, but cannot be reached from any of the preceding sources $s_1, s_2, \dots, s_{i - 1}$. Note that $\sum_{i=1}^{k} |\comp_\malloc(s_i)| = n$.

The fourth updating rule applies when the first three rules are not applicable, and there are $l \ge 1$ distinct goods $\good_0, \good_1, \dots, \good_{l - 1}$ in $\malloc_0$, $l$ distinct sources $s'_0, s'_1, \dots s'_{l - 1}$, and $l$ distinct agents $t_1, t_2, \dots t_l$ such that $t_{i\%l} \in \comp_\malloc(s'_i)$ and $t_{i + 1}$ is the most envious agent of $\malloc_{s'_i} \cup \{\good_i\}$ for $0 \le i < l$. Then we shift the bundles of agents through cycle $\cycle = s'_0,\ldots,t_1,s'_1,\ldots,t_2,s'_2,\ldots,t_{l},s'_0$, i.e, we set:
\begin{align*}
	&\malloc'_{t_{i + 1}} \leftarrow \mbox{The smallest subset of $\malloc_{s'_i} \cup \{\good_i\}$ which is envied by $t_{i+1}$}.\\
	&\malloc'_i \leftarrow \malloc_j \: \mbox{For every edge $i \rightarrow j$ in the cycle.}
\end{align*}

\begin{lemma}\label{lem:751}
	Let $\malloc$ be an $\efx$ allocation satisfying Property \ref{prop:51}. Assume that Rule 1, Rule 2, and Rule 3 are not applicable on $\malloc$, and let $\malloc'$ be the allocation after applying Rule 4 on $\malloc$.
	Then $\malloc'$ is $\efx$ and satisfies Property \ref{prop:51}. Furthermore, we have $\phi(\malloc') > \phi(\malloc)$. 
\end{lemma}
\begin{proof}
	For each non-source agent $i$ in $G_{1, \malloc'}$ we need to prove $\malloc'_i = \relevant{\malloc'_i}{i}$. By the way we shift the bundles through the cycle, for every agent in the cycle, the statement holds trivially. For every $j \notin \cycle$, we know that the bundle of this agent has not changed. Also, any agent $i \notin C$ that was a source prior to the update, remains a source thereafter. Hence, Property \ref{prop:51} holds after the update.
	In addition, after this update, no agent is worse-off. Moreover agents in the cycle are strictly better-off. Therefore, $\phi(\malloc') > \phi(\malloc)$.
\end{proof}

We also prove  Lemma \ref{q1-goods-bound} on the applicability of Rule 4. We later use this lemma to prove an upper bound on the number of remaining goods, when none of Rules $1-4$ are applicable. 
\begin{lemma} \label{q1-goods-bound}
	Let $\malloc$ be an $\efx$ allocation satisfying Property \ref{prop:51}. Assume that Rule 1, Rule 2, and Rule 3 are not applicable on $\malloc$.
	Let $s'_1, s'_2, \dots, s'_l$ be subset of sources of $G_{1,\malloc}$ such that for every $1\leq i \leq l$, $|\comp_\malloc(s'_i)|\geq 2$. If $\malloc_0$ contains at least $l$ goods, then Rule 4 is always applicable.
\end{lemma}
\begin{proof}
	Assume $\good_1, \good_2, \dots \good_l$ are $l$ goods in $\malloc_{0}$. For each good $\good_i$, let $t_i$ be the most envious agent of the set of goods $\malloc_{s'_i} \cup \{\good_i\}$. We construct a new graph with $l$ vertices, where each vertex represents one of the sources. For each $1 \le i \le l$, we add a directed edge from $s'_i$ to the source $s'_j$ where $t_i \in s'_j$. Therefore, the new graph has $l$ vertices and every vertex has exactly one outgoing edge. Hence, this graph admits a cycle which corresponds to a situation satisfying the conditions of Rule 4. 
\end{proof}

\subsection{Final Step}

When no updating rule is applicable, we head to a final step to finalize the allocation.
However, before describing our final step, we show that when none of the rules $1-4$ is feasible, we are guaranteed that $|\malloc_0|< \frac{n}{2}$, which means that $\malloc$ is an $\efx$ allocation with less than $\frac{n}{2}$ number of discarded goods. 

\begin{theorem}\label{thm7}
	For every instance with $(\infty,1)$-bounded valuations, there exists an $\efx$ allocation which discards no more than $\lfloor \frac{n}{2} \rfloor - 1$ goods.
\end{theorem}
\begin{proof}
	By Lemma \ref{q1-goods-bound}, when none of Rules 1-4 is applicable, the number of remaining goods in the pool is strictly less than the number of sources whose component contains at least $2$ agents. Given that $\sum_{i=1}^{k} |\comp_\malloc(s_i)| = n$, there are at most $\lfloor \frac{n}{2} \rfloor$ such sources. Consequently, the number of remaining goods in the pool is at most $\lfloor \frac{n}{2} \rfloor - 1$.
\end{proof}

The purpose of the final step is to allocate all the remaining goods in the pool and achieve a complete $\eftx$ allocation.
Let $\good_1, \good_2, \dots \good_x$ denote the remaining goods in the pool, and let $s'_1, s'_2, \dots, s'_y$ be the sources with at least $2$ agents in their components. Since Rule 4 is not applicable, by Lemma \ref{q1-goods-bound} we have  $x<y$. In the final step, for every $1 \le i \le x$, we allocate $\good_i$ to an arbitrary non-source agent in $\comp_\malloc(s'_i)$.

\begin{theorem}\label{thm7f}
	Suppose the valuations are $(\infty,1)$-bounded and let $\malloc^*$ be the allocation returned by $\algo$.
Then, $\malloc^*$ is $\eftx$.
\end{theorem}
\begin{proof}
Let $\malloc$ be the allocation before the final step.
By Theorem \ref{thm7} we know that $\malloc$ is $\efx$.
Let $j$ be an agent that receives a good in the final step. 
We claim that before the final step, $|\malloc_j|=1$. To show this, note that since $j$ is not a source, before the final step, some agent $i$ envies $\malloc_j$. However, since the allocation is $\efx$, agent $i$ does not envy any strict subset of $\malloc_j$, which means $\relevant{\malloc_j}{i} = \malloc_j $. However, by definition of $(\infty,1)$-bounded valuations, there can be at most one good with a non-zero value to both agents $i$ and $j$. Hence, $|\malloc_j|=1$. Consequently, the size of $\malloc^*_j$ after the final step will be $2$. Therefore, the allocation is envy-free up to any two goods.
\end{proof}
	\section{$\frac{\sqrt{2}}{2}$-$\efx$ for Restricted Additive Valuations}\label{sec:5}
In this section, we present a $\frac{\sqrt{2}}{2}$-$\efx$ allocation algorithm for restricted additive valuations. The algorithm uses two updating rules and a final step. After each update, an envy elimination process ensures the allocation remains $\invsq$-$\efx$. Our algorithm also guarantees that Property \ref{prop:2} holds at every step before the final step. This property holds trivially for the basic feasible allocation.

\begin{enumerate}[label=\textcolor{red}{$\S\!\!$} (\roman*)]
	\item\!\!: \label{prop:2}	For every agent $i$, $\relevant{\malloc_i}{i} = \malloc_i$. 	
\end{enumerate}



Given Lemma \ref{lem:11}, Property \ref{prop:2} implies that $G_{\sqrt{2},\malloc}$ is always acyclic.  \footnote{Note that this also holds for the initial allocation.}  We assume that $G_{\sqrt{2},\malloc}$ contains $k$ source agents. For convenience, we also assume that the agents are re-indexed after each update so that agents $1,2,\ldots,k$ are the sources of $G_{\sqrt{2},\malloc}$, and $\valu(\malloc_1) \leq \valu(\malloc_2) \leq \ldots \leq \valu(\malloc_k)$. We define our potential function as 
$
\phi(\malloc) = \big[ \valu(\malloc_1),\valu(\malloc_2),\ldots, \valu(\malloc_k),\infty \big].
$ We designate our algorithm in this section as $\algot$. 

\subsection{Envy Elimination} \label
{sec:ee}
As we mentioned, the envy elimination process is initially introduced by Akrami $\etal$ \cite{akrami2022ef2x}. This process aims to eliminate strong envies among agents while preserving essential allocation properties. Here, we propose a simple alternative for the envy elimination process: Given an allocation $\malloc$ with potential strong envies, as long as there exist two agents, $i$ and $j$, where agent $i$ $\sqrt{2}$-strongly envies agent $j$, we find a good $\good \in \malloc_j$ such that $\sqrt{2}\valu_i(\malloc_i) < \valu_i(\malloc_j \setminus \{\good\})$, and move $\good$ from $\malloc_j$ to $\malloc_0$. 

 Given the nature of the process, we know that the resulting allocation is $\invsq$-$\efx$. As we show in Lemma \ref{lem:2}, this process also guarantees that the value of the potential function does not decrease.

\begin{lemmarep}\label{lem:2}
	Let $\malloc$ be a $\frac{\sqrt{2}}{2}$-$\efx$ allocation satisfying Property \ref{prop:2}, and let $\malloc'$ be the allocation after applying envy elimination. Then, $\malloc'$ satisfies Property \ref{prop:2}. Furthermore we have $\phi(\malloc) \leq \phi(\malloc')$.  
\end{lemmarep}
\begin{proof}
	By definition of the process, Property \ref{prop:2}  holds trivially. To show that $\phi(\malloc) \leq \phi(\malloc')$, note that the envy elimination process may involve multiple updates. We show that after each update, the value of $\phi$ does not decrease. Consider an allocation $\malloc$ where, for some agents $i$ and $j$, we have $\sqrt{2}\valu_i(\malloc_i) < \valu_i({\malloc_j / \{\good\}})$. Let $\malloc''$ be the allocation after the update, which means $\malloc''$ is identical to $\malloc$ except that good $\good$ is transferred from the bundle of agent $j$ to the pool. Note that agent $j$ is neither a source before the update nor becomes a source after, since agent $i$ $\sqrt{2}$-envies agent $j$ both before and after the process. The only change in the source nodes happens when agent $j$ starts to $\sqrt{2}$-envy some source node after the process. Therefore, the set of sources in $G_{\sqrt{2}, \malloc}$ is a subset of the set of sources $G_{\sqrt{2}, \malloc''}$. Therefore, by definition of $\phi(\cdot)$, we have $\phi(\malloc) \leq \phi(\malloc'')$, and  consequently, $\phi(\malloc) \leq \phi(\malloc')$.
\end{proof}


\subsection{Update Rule 1} \label{aa:1}
The first rule takes an allocation $\malloc$ as input, and if there exists a source agent $i$ and a good $\good \in \malloc_0$ such that good $\good$ is relevant to agent $i$,  we allocate $\good$ to agent $i$. 

\begin{lemmarep}\label{lem:511}
	Let $\malloc$ be an allocation satisfying Property \ref{prop:2}, and let $\malloc'$ be the allocation after applying Rule 1. Then $\malloc'$ satisfies Property \ref{prop:2}. Furthermore, we have $\phi(\malloc) < \phi(\malloc')$.
\end{lemmarep}
\begin{proof} 
	The only agent whose value for her bundle is changed is agent $i$, and the allocated good $\good$ is relevant to agent $i$. Therefore, Property \ref{prop:2} holds. Furthermore, $\valu(\malloc'_i) > \valu(\malloc_i)$ and  by Lemma \ref{lem:phi} we have $\phi(\malloc) < \phi(\malloc')$. 
\end{proof}

\subsection{Update Rule 2} \label{aa:2}
In the second updating rule, if for some agent $i$,  $\valu_i(\malloc_i) < \sqrt{2}\valu_i(\malloc_0)$, we update the allocation as follows: since Rule 1 is not applicable,  $i$ is not a source. Also, since $G_{\sqrt{2},\malloc}$ is acyclic, there is a source, let's say $j$,  with a directed path to $i$ in $G_{\sqrt{2},\malloc}$. We shift the allocation through this path, i.e., for every edge $i_1 \rightarrow i_2$ in this path, we allocate $\malloc_{i_2}$ to agent $i_1$. We also give all the goods that are valuable to agent $i$ in $\malloc_0$ to $i$, and return the goods in $\malloc_j$ back to the pool. 

\begin{lemmarep}\label{lem:521}
Let $\malloc$ be an allocation satisfying Property \ref{prop:2}. Assume that Rule 1 is not applicable,  and let $\malloc'$ be the allocation after applying Rule 2. Then, $\malloc'$ satisfies Property \ref{prop:2}. Furthermore, we have $\phi(\malloc) < \phi(\malloc')$.
\end{lemmarep}
\begin{proof}
	Given that agent $i$ selects only relevant goods from the pool, we have $\valu_i(\malloc'_i) = \valu({\malloc'_i})$. Furthermore, according to Lemma \ref{lem:12}, for every edge $i_1 \rightarrow i_2$ in the path from $j$ to $i$, we have  $\valu_{i_1}(\malloc'_{i_2}) = \valu(\malloc'_{i_2})$. Consequently, Property \ref{prop:2} remains valid after the update.
	To prove $\phi(\malloc) < \phi(\malloc')$, we utilize Lemma \ref{lem:phi}. Considering $A,x,x'$ in the statement of Lemma \ref{lem:phi} we proceed with: 
	\begin{align*}
		A &= \mbox{Agents in the path from $j$ to $i$ } \\
		x &= \min_{i' \in A}(\valu(\alloc_{i'})) = \valu(\malloc_j) \\
		x' &= \min_{i' \in A}(\valu(\alloc'_{i'}))
	\end{align*}
	Also, since for agent $i$, $\valu(\malloc_j) < \invsq\valu(\malloc_i) 
	< \valu(\malloc'_i),
	$
	and for every other agent $k \in A$, $\valu(\malloc_j) < \invsq\valu(\malloc_{k})$, we have $x < x'$.  
	Finally, agent $j$ is a source in $G_{\sqrt{2}, \malloc}$ and therefore, $x = \valu(\malloc_j)$. Thus, all conditions of Lemma \ref{lem:phi} are satisfied, which implies $\phi(\malloc) < \phi(\malloc')$.
\end{proof}

\subsection{Final Step} 
When none of Rules 1 and 2 is feasible, we give the entire pool to agent $1$ and terminate the allocation. 
In Theorem \ref{thm:fstep} we show  that the resulting allocation is $\frac{\sqrt{2}}{2}$-$\efx$. 

\begin{theoremrep}\label{thm:fstep}
	Suppose the valuations are restricted additive, and let $\malloc^*$ be the allocation returned by $\algot$. Then, $\malloc^*$ is a $\frac{\sqrt{2}}{2}$-$\efx$ allocation.
\end{theoremrep}
\begin{proof}
	Let $\malloc$ be the allocation before the final step.
	First, note that agent $1$ is a source in $G_{\sqrt{2}, \malloc}$. We prove that after this allocation, no-one will $\sqrt{2}$-envy $\malloc^*_1$. As a contradiction, 
	suppose agent $i$ $\sqrt{2}$-envies agent $1$. Since agent $1$ was a source, agent $i$ did not $\sqrt{2}$-envy him before the final step. This implies that the pool has a non-zero value for agent $i$, and given the fact that Rule 1 was not applicable, agent $i$ is not a source. Now, suppose that agent $j$ is a source that has a path to $i$. We have
	\begin{align*}
		\valu_i(\malloc^*_1) &= \valu(\malloc_1) + \valu_i(\malloc_0) \\
		&\leq  \valu(\malloc_j) + \valu_i(\malloc_0) &\valu(\malloc_1) \leq\valu(\malloc_j)  \\
		&\leq \frac{\sqrt{2}}{2}\valu(\malloc_i)  + \valu_i(\malloc_0) & \mbox{ $j$  has a path to $i$ in }G_{\sqrt{2},\malloc}\\
		&\leq \frac{\sqrt{2}}{2}\valu(\malloc_i)  + \frac{\sqrt{2}}{2}\valu(\malloc_i) & \mbox{$i$ does not $\sqrt{2}$-envy $\malloc_0$}\\
		&\leq \sqrt{2}\valu(\malloc_i). 
	\end{align*}
	But this contradicts the fact that agent $i$ does not $\sqrt{2}$-envy agent $j$ before the final step.
\end{proof}

\label{case1}

	\section{$\frac{\sqrt{2}}{2}$-$\efx$ for $(\infty,1)$-bounded Valuations} \label{sec:eh}

In this section, we propose a $\invsq$-$\efx$ allocation algorithm for agents with $(\infty,1)$-bounded subadditive valuations. Our algorithm consists of a single updating rule, followed by a final step. 
Throughout the algorithm, we use the Nash social welfare of the allocation as our potential function, that is, $\phi(\malloc) = \nsw(\malloc)$. We also ensure that the following properties are consistently maintained throughout the algorithm before the final step:

\begin{enumerate}[label=\textcolor{red}{$\dag\!\!$} (\roman*)]
	\item \label{q1-no-greater-1-cycle} LP \eqref{lp1} is feasible for allocation $\malloc$.
	\item \label{q1-virtually-efx} No agent \textbf{virtually strongly envies} another agent. Formally, for any pair of agents $i$ and $j$, and any good $g \in \malloc_j$, we have:
	$\virtualvalue_i(\malloc) \geq v_i(\malloc_j \setminus {g})$.
	\item \label{q1-relevant} For every agent $i$, $\relevant{\malloc_i}{i} = \malloc_i$.
\end{enumerate}

We designate our algorithm in this section as $\algoth$. 
\subsection{Update Rule 1} \label{general-rank-pool-rule}
This update rule is applicable when an agent virtually envies the pool. Let $\hat{\malloc_0} \subseteq \malloc_0$ denote a minimal subset of the pool that is virtually envied by at least one agent. Consider agent $i$ with the maximum value of $v_i(\hat{\malloc_0})/\virtualvalue_i(\malloc)$. 
Note that all goods in $\hat{\malloc_0}$ are relevant to agent $i$. Let $P$ denote the rankpath of agent $i$. We allocate $\hat{\malloc_0}$ to agent $i$ and shift the bundles along path $P$. Formally, for every edge $j \rightarrow k \in P$, we allocate $\malloc_k$ to agent $j$. Then, we return the bundle of the first agent in $P$ to the pool. Note that this update rule does not rely on properties of $(\infty, 1)$-bounded valuations and can be used in the general setting. Thus, we state our lemmas in this section in general form.

To prove that $\malloc'$ satisfies the required properties, we define $\rupp_j$ for each agent $j$ as follows:

\begin{align} \label{rank-upperbound-def}
	\rupp_j = \begin{cases}
		\rank_j(\malloc), & \text{if } j \notin P \\
		\rank_k(\malloc), & \text{if } j \rightarrow k \in P \\
		\frac{v_j(X_0^*)}{\virtualvalue_j(\malloc)}, & \text{if } j = i
	\end{cases}.
\end{align}

We now show that $\rupp$ is a feasible solution to LP \eqref{lp1} for allocation $\malloc'$.

\begin{lemmarep} \label{same-virtual-value}
	Let $\malloc$ be an allocation satisfying Properties \ref{q1-no-greater-1-cycle} and \ref{q1-virtually-efx}, and let $\malloc'$ be the allocation after applying Rule 1. Then, for every agent $j$, $\frac{\valu_j(\malloc'_j)}{\rupp_j} = \virtualvalue_j(\malloc).$
\end{lemmarep}
\begin{proof} 
	We distinguish among three cases:
	
	\textbf{Case 1 ($j \notin P$):} We have $\rupp_j = \rank_i(\malloc)$ and $\malloc'_j = \malloc_j$. This implies: $$\virtualvalue_j(\malloc) = \frac{\valu_j(\malloc_j)}{\rank_i(\malloc)} = \frac{\valu_j(\malloc'_j)}{\rupp_j}.$$
	
	\textbf{Case 2 ($j \rightarrow k \in P$):} After shifting the bundles through $P$, agent $j$ receives the bundle of agent $k$. Furthermore, since $P$ is the rankpath of agent $i$, by Observation \ref{obs:rankpath} we have $\rank_k(\malloc) = \rank_j(\malloc)  \frac{\valu_j(\malloc_k)}{\valu_j(\malloc_j)}$. Hence:
	\begin{align*}
		\frac{\valu_j(\malloc'_j)}{\rupp_j} &= \frac{\valu_j(\malloc_k)}{\rank_k(\malloc)} & \mbox{$\malloc'_j = \malloc_k$ and $\rupp_j = \rank_k(\malloc)$} \\
		&= \frac{\valu_j(\malloc_j)}{\rank_j(\malloc)} = \virtualvalue_j(\malloc) & \mbox{Observation \ref{obs:rankpath}}.
	\end{align*}
	
	\textbf{Case 3 ($j = i$):} Since $\malloc'_i = \hat{\malloc_0}$, we have:
	\begin{align*}
		\frac{\valu_i(\malloc'_i)}{\rupp_i} &= \frac{\valu_i(\hat{\malloc_0})}{\rupp_i} & \mbox{$\malloc'_i = \hat{\malloc_0}$} \\
		&= \frac{\valu_i(\hat{\malloc_0})}{\frac{\valu_i(\hat{\malloc_0})}{\virtualvalue_i(\malloc)}} = \virtualvalue_i(\malloc) & \mbox{$\rupp_i = \frac{\valu_i(\hat{\malloc_0})}{\virtualvalue_i(\malloc)}$}.
	\end{align*}
\end{proof}

\begin{lemmarep} \label{rupp-satisfies-rank-property}
	Let $\malloc$ be an allocation satisfying Properties \ref{q1-no-greater-1-cycle} and \ref{q1-virtually-efx}, and let $\malloc'$ be the allocation after applying Rule 1. Then, $\rupp$ is a feasible solution to LP \eqref{lp1} for allocation $\malloc'$.
\end{lemmarep}
\begin{proof} 
	In order to show that for every agents $j$ and $k$, $\rupp_j w_{\malloc'}(j, k) \le \rupp_k$, we distinguish three cases:
	
	\textbf{Case 1 ($k \notin P$):} The bundle of agent $k$ remains the same as it was before. Therefore:
	\begin{align*}
		\rupp_j w_{\malloc'}(j, k) &= \frac{\rupp_j}{\valu_j(\malloc'_j)} \valu_j(\malloc'_k) & \mbox{Definition \ref{weighted-envy-def}} \\
		&= \frac{\valu_j(\malloc'_k)}{\virtualvalue_j(\malloc)} & \mbox{Lemma \ref{same-virtual-value}} \\
		&= \frac{\valu_j(\malloc_k)}{\virtualvalue_j(\malloc)} & \mbox{$\malloc'_k = \malloc_k$} \\
		&= \rank_j(\malloc) \frac{\valu_j(\malloc_k)}{\valu_j(\malloc_j)} & \mbox{Definition \ref{virtual-value-def}}\\
		&\le \rank_k(\malloc) & \mbox{ Property \ref{q1-no-greater-1-cycle}} 
		\\
		&= \rupp_k. & \mbox{Equation \eqref{rank-upperbound-def}}
	\end{align*}
	
	\textbf{Case 2 ($k \rightarrow l \in P$):} Agent $k$ receives the bundle of agent $l$. Additionally, by Equation \eqref{rank-upperbound-def}, we have $\rupp_k = \rank_l(\malloc)$. Therefore:
	
	\begin{align*}
		\rupp_j w_{\malloc'}(j, k) &= \frac{\rupp_j}{\valu_j(\malloc'_j)} \valu_j(\malloc'_k) & \mbox{Definition \ref{weighted-envy-def}} \\
		&= \frac{\valu_j(\malloc'_k)}{\virtualvalue_j(\malloc)} & \mbox{Lemma \ref{same-virtual-value}} \\
		&= \frac{\valu_j(\malloc_l)}{\virtualvalue_j(\malloc)} & \mbox{$\malloc'_k = \malloc_l$} \\
		&= \rank_j(\malloc) \frac{\valu_j(\malloc_l)}{\valu_j(\malloc_j)} & \mbox{Definition \ref{virtual-value-def}}\\
		&\le \rank_l(\malloc) & \mbox{$\malloc$ satisfies Property \ref{q1-no-greater-1-cycle}} 
		\\
		&= \rupp_k & \mbox{Equation \eqref{rank-upperbound-def}}.
	\end{align*}
	
	\textbf{Case 3 ($k = i$):} Agent $i$ receives $\hat{\malloc_0}$. Note that since $i$ has the maximum value of $\frac{v_i(\hat{\malloc_0})}{\virtualvalue_i(\malloc)}$ we have $\frac{v_j(\hat{\malloc_0})}{\virtualvalue_j(\malloc)} \le \frac{v_i(\hat{\malloc_0})}{\virtualvalue_i(\malloc)}$. This implies:
	
	\begin{align*}
		\rupp_j w_{\malloc'}(j, i) &= \frac{\rupp_j}{\valu_j(\malloc'_j)} \valu_j(\malloc'_i) & \mbox{Definition \ref{weighted-envy-def}} \\
		&= \frac{\valu_j(\malloc'_i)}{\virtualvalue_j(\malloc)} & \mbox{Lemma \ref{same-virtual-value}} \\
		&= \frac{\valu_j(\hat{\malloc_0})}{\virtualvalue_j(\malloc)} & \mbox{$\malloc'_i = \hat{\malloc_0}$} \\
		&\le \frac{v_i(\hat{\malloc_0})}{\virtualvalue_i(\malloc)} &\mbox{ $\frac{v_i(\hat{\malloc_0})}{\virtualvalue_i(\malloc)}$ is maximum} \\
		&= \rupp_i & \mbox{Equation \eqref{rank-upperbound-def}}.
	\end{align*}
\end{proof}

\begin{lemmarep} \label{virtual-value-increases}
	Let $\malloc$ be an allocation satisfying Properties \ref{q1-no-greater-1-cycle} and \ref{q1-virtually-efx}, and let $\malloc'$ be the allocation after applying Rule 1. Then for every agent $i$ we have $\virtualvalue_j(\malloc) \le \virtualvalue_j(\malloc')$.
\end{lemmarep}
\begin{proof} 
	Lemma \ref{rupp-satisfies-rank-property} ensures that for every agents $j$ and $k$ we have $\rupp_j w_{\malloc'}(j, k) \le \rupp_k$. According to Lemma \ref{upper-bound-no-cycle}, this guarantees $G_{\malloc'}$ does not have any cycle whose total product of the edge weights is strictly greater than $1$. Additionally, by Lemma \ref{rank-upper-bound}, for each agent $j$, we have $\rank_j(\malloc') \le \rupp_j$ and Lemma \ref{same-virtual-value} implies that $\virtualvalue_j(\malloc) = \frac{\valu_j(\malloc'_j)}{\rupp_j}$. Consequently, we have $\virtualvalue_j(\malloc') = \frac{\valu_j(\malloc'_j)}{\rank_j(\malloc')} \ge \frac{\valu_j(\malloc'_j)}{\rupp_j} = \virtualvalue_j(\malloc)$.
\end{proof}

\begin{lemmarep} \label{no-virtual-se}
	Let $\malloc$ be an allocation satisfying Properties \ref{q1-no-greater-1-cycle} and \ref{q1-virtually-efx}, and let $\malloc'$ be the allocation after applying Rule 1. Then, for every agents $j$ and $k$ and good $g \in \malloc'_k$, we have $\virtualvalue_j(\malloc') \ge \valu_j(\malloc'_k \setminus \{g\})$.
\end{lemmarep}
\begin{proof} 
	We show that every agent $j$ does not virtually strongly envy another agent $k$. If $k \ne i$, there exists a (not necessarily different from $k$) agent $l$ such that $\malloc'_k = \malloc_l$. Therefore, for every good $g \in \malloc'_k = \malloc_l$ we have:
	\begin{align*}
		\virtualvalue_j(\malloc') &\ge \virtualvalue_j(\malloc) & \mbox{Lemma \ref{virtual-value-increases}} \\
		&\ge \valu_j(\malloc_l \setminus \{g\}) & \mbox{$\malloc$ satisfies Property \ref{q1-virtually-efx}} \\
		&= \valu_j(\malloc'_k \setminus \{g\}) & \mbox{$\malloc'_k = \malloc_l$}.
	\end{align*}
	Additionally, for agent $i$, we know that $i$ receives $\hat{\malloc_0}$, which is the minimal subset that is virtually envied by at least one agent. Hence, $j$ does not virtually strongly envy agent $i$.
\end{proof}

\begin{lemmarep} \label{lem:q1-relevant-goods}
	Let $\malloc$ be an allocation satisfying Properties \ref{q1-no-greater-1-cycle} and \ref{q1-virtually-efx} and \ref{q1-relevant}, and let $\malloc'$ be the allocation after applying Rule 1. Then, for every agent $j$, $\malloc'_j[j] = \malloc'_j$.
\end{lemmarep}
\begin{proof}
	If $j \notin P$, since $\malloc'_j = \malloc_j$, the property holds trivially. Additionally, since $\hat{\malloc_0}$ is minimal, all the goods in $\malloc'_i$ are relevant to agent $i$. Now, we only need to show that for every agent $j \rightarrow k \in P$, all the goods in $\malloc'_j = \malloc_k$ are relevant to agent $j$. Assume by contradiction that there exists a good $g \in \malloc_k = \malloc'_j$ that is not relevant to agent $j$. Then, we have:
	\begin{align*}
		\valu_j(\malloc_k \setminus \{g\}) &= \valu_j(\malloc_k) & \mbox{$g$ is not relevant to $j$} \\
		&= \valu_j(\malloc_j) w_\malloc(j, k) & \mbox{Definition \ref{weighted-envy-def}} \\
		&= \valu_j(\malloc_j) \frac{\rank_k(\malloc)}{\rank_j(\malloc)} & \mbox{Observation \ref{obs:rankpath}} \\
		&= \virtualvalue_j(\malloc) \rank_k(\malloc) & \mbox{Definition \ref{virtual-value-def}} \\
		&> \virtualvalue_j(\malloc) & \mbox{$\rank_k(\malloc) > 1$}.
	\end{align*}
	Therefore, agent $j$ virtually strongly envies $k$ in allocation $\malloc$. This contradicts the fact that $\malloc$ satisfies Property \ref{q1-virtually-efx}.
\end{proof}

\begin{lemmarep} \label{nsw-increases}
	Let $\malloc$ be an allocation satisfying Properties \ref{q1-no-greater-1-cycle} and \ref{q1-virtually-efx}, and let $\malloc'$ be the allocation after applying Rule 1. Then, $\nsw(\malloc') > \nsw(\malloc)$.
\end{lemmarep}
\begin{proof}
	Since the bundle of agents that are not in $P$ is still the same as before, it is sufficient to show that the product of valuation of agents in path $P$ increases.
	\begin{align*}
		\prod_{j \in P} \valu_j(\malloc'_j) &= \valu_i(\hat{\malloc_0}) \prod_{j \rightarrow k \in P} \valu_j(\malloc_k) & \mbox{Bundles are shifted through path $P$} \\
		&> \virtualvalue_i(\malloc) \prod_{j \rightarrow k \in P} \valu_j(\malloc_k) &\mbox{$\valu_i(\hat{\malloc_0}) > \virtualvalue_i(\malloc)$} \\
		&= \frac{\valu_i(\malloc_i)}{\rank_i(\malloc)} \prod_{j \rightarrow k \in P} \valu_j(\malloc_k) &\mbox{Definition \ref{virtual-value-def}} \\
		&= \frac{\valu_i(\malloc_i)}{\rank_i(\malloc)} \prod_{j \rightarrow k \in P} \valu_j(\malloc_j) w_\malloc(j, k) &\mbox{Definition \ref{weighted-envy-def}} \\
		&= \frac{\rank_i(\malloc)}{\rank_i(\malloc)} \prod_{j \in P} \valu_j(\malloc_j) & \mbox{$P$ is rankpath of $i$} \\
		&= \prod_{j \in P} \valu_j(\malloc_j).
	\end{align*}
	Therefore $\nsw(\malloc') > \nsw(\malloc)$.
\end{proof}

In conclusion, after applying this update rule, the resulting allocation is virtually $\efx$, virtual values of agents do not decrease, and Nash social welfare of the allocation strictly increases.

\begin{corollary} [of Lemmas \ref{rupp-satisfies-rank-property}, \ref{no-virtual-se}, \ref{lem:q1-relevant-goods} and \ref{nsw-increases}]
	Let $\malloc$ be an allocation satisfying Properties \ref{q1-no-greater-1-cycle} and \ref{q1-virtually-efx} and \ref{q1-relevant}, and let $\malloc'$ be the allocation after applying Rule 1. Then, $\malloc'$ satisfies Properties \ref{q1-no-greater-1-cycle},  \ref{q1-virtually-efx}, and \ref{q1-relevant}. Furthermore, we have $\phi(\malloc') > \phi(\malloc)$.
\end{corollary}	
\subsection{Final Step}
When Rule 1 is not applicable, we perform a final step to achieve a complete $\invsq$-$\efx$ allocation. Let $i$ be an agent with rank equal to one in $\malloc$, and $\sagents$ be the set of all other agents whose rank is at most $\sqrt{2}$. In the final step, first we iterate through $\sagents$ in an arbitrary order and ask each agent to pick all the goods in the pool that are relevant to her. Then we allocate all the remaining goods in the pool to agent $i$ and return the allocation.

\begin{theoremrep}\label{thm72}
	Suppose the valuations are $(\infty, 1)$-bounded, and let $\malloc^*$ be the allocation returned by $\algoth$. Then, $\malloc^*$ is a $\frac{\sqrt{2}}{2}$-$\efx$ allocation.
\end{theoremrep}
\begin{proof}
	First we show that for every agents $j \in \sagents$ and $k \in \agents$, agent $k$ does not $\sqrt{2}$-envy agent $j$. Note that since $\malloc_k \subseteq \malloc^*_k$ we have $\valu_k(\malloc_k) \le \valu_k(\malloc^*_k)$. Additionally, since agent $j$ only receives the goods from the pool that are relevant to her, and the valuations are $(\infty, 1)$-bounded, at most one good in $\malloc^*_j$ is relevant to agent $k$. We denote this good by $g$. If $g \in \malloc_0$, since Rule 1 is not applicable $\valu_k(\malloc_k) \ge \virtualvalue_k(\malloc) \le \valu_k(g)$ and $k$ does not envy agent $j$; otherwise $g \in \malloc_j$, and since $\rank_j(\malloc) \le \sqrt{2}$ agent $k$ does not $\sqrt{2}$-envy $\malloc_j$. Hence, agent $k$ does not $\sqrt{2}$-envy agent $j$.
	
	Now, we need to prove that there is no agent $k \in \agents$ who $\sqrt{2}$-envies $\malloc^*_i$. In the case that $k \in \sagents$, since $k$ has picked all the goods from the pool which are relevant to her, agent $i$ does not receive any good that is relevant to agent $k$. Furthermore, since $\rank_i(\malloc) = 1$, agent $k$ does not envy agent $i$. For the case that $\rank_k(\malloc) > \sqrt{2}$, since Rule 1 is not applicable we have $\valu_k(\malloc_0) < \virtualvalue_k(\malloc) \le \invsq \valu_k(\malloc_k)$. Additionally, since $\malloc$ satisfies Property \ref{q1-no-greater-1-cycle} we have $\rank_k(\malloc) w_{\scriptscriptstyle \malloc}(k, i) \le \rank_i(\malloc)$, and since $\rank_i(\malloc) = 1$ and $\rank_k(\malloc) > \sqrt{2}$ we have $\valu_k(\malloc_i) \le \invsq \valu_k(\malloc_k)$. Therefore:
	\begin{align*}
		\valu_k(\malloc^*_i) &\le \valu_k(\malloc_i \cup \malloc_0) &\mbox{$\malloc^*_i \subseteq \malloc_i \cup \malloc_0$} \\
		&\le \valu_k(\malloc_i) + \valu_k(\malloc_0) & \mbox{Subadditivity} \\
		&\le \valu_k(\malloc_i) + \invsq \valu_k(\malloc_k) & \mbox{Rule 1 is not applicable} \\
		&\le \invsq \valu_k(\malloc_k) + \invsq \valu_k(\malloc_k) & \mbox{$\rank_i(\malloc) = 1$} \\
		&= \sqrt{2} \valu_k(\malloc_k).
	\end{align*}
	Hence, there is no agent $k$ that $\sqrt{2}$-envies agent $i$.
	
	It only remains to prove that for every agents $j$ and $k$ such that $\rank_k > \sqrt{2}$, agent $j$ does not strongly-envy agent $k$. Note that $\malloc$ satisfies Property \ref{q1-virtually-efx}, which means for every good $g \in \malloc_k$ we have $\virtualvalue_j(\malloc) \ge \valu_j(\malloc_k \setminus \{g\})$. Additionally, since $\rank_k(\malloc) > \sqrt{2}$, $k$ does not receive any good in the final step, i.e., $\malloc^*_k = \malloc_k$. Therefore $\valu_j(\malloc^*_j) \ge \virtualvalue_j(\malloc) \ge \valu_j(\malloc_k \setminus \{g\})$ and agent $j$ does not strongly envy agent $k$. 
	This completes the proof.
\end{proof}

\label{case2}
	\section{$\efx$ for the Restricted Additive Valuations with p = 2}\label{sec:8}

Our final algorithm finds a complete $\efx$ allocation for $(2,\infty)$-bounded and restricted additive valuations. With $p=2$, each good is relevant to at most two agents, while any pair of agents may share multiple relevant goods. Here, for a set $S$, $\relevant{S}{i,i}$ refers to goods relevant only to agent $i$.

Our algorithm consists of four updating rules and a final step. Unlike our previous algorithms, agents may become satisfied and be eliminated from the process. Therefore, here $\agents$ refers to the set of remaining agents. We also define rank, rankpath, and virtual value based on the allocation of the remaining agents. We use $ \phi(\malloc) = (n-|\agents|,\nsw(\malloc))$ as the potential function to prove the termination of our algorithm:  Therefore, after each update, either the number of satisfied agents increases or the Nash social welfare of the allocation increases. We designate our algorithm in this section as $\algopr$. 
 Furthermore, our updates are designed in a way that the following properties are always satisfied during the algorithm:

\begin{enumerate}[label=\textcolor{red}{$\uparrow\!\!$} (\roman*)]
	\item \label{prop42}
For every agent $i \in \agents$, there is an agent $j$ such that $\malloc_i \subseteq \relevant{\goods}{i,j}$. We call agent $j$, the corresponding agent of agent $i$. Note that the corresponding agent of $i$ can be $i$ itself.
\item \label{prop43} For every agent $i,j \in \agents$, and every 
$
\good \in \malloc_j$ we have $\valu_i(\malloc_j \setminus \{\good\}) \le \virtualvalue_i(\malloc)$.  
\end{enumerate}
Note that Properties \ref{prop42} and \ref{prop43} hold  trivially for the basic feasible allocation. We also prove that after each update, the virtual value of each agent does not decrease. This property also plays a key role in our proofs.


\begin{observationrep}\label{obs41}
	Let $\alloc$ be an allocation that satisfies Property \ref{prop42}. Then for every $i \in \agents$, we have $\virtualvalue_i(\alloc) = \valu(\alloc_{\roott{i}{\alloc}})$.
\end{observationrep}
\begin{proof} 
	Let $i_1 \rightarrow i_2 \rightarrow \dots \rightarrow i_k \rightarrow i$ be the rankpath of agent $i$ in $G_{\alloc}$. 
	
	\begin{align*}
		\rank_i(\alloc) &= w_{\scriptscriptstyle \alloc}(i_1, i_2) \times w_{\scriptscriptstyle \alloc}(i_2, i_3) \times \ldots \times w_{\scriptscriptstyle \alloc}(i_k, i) &\text{Definition \ref{weighted-envy-def}} \\
		&= \frac{\valu_{i_1}(\alloc_{i_2})}{\valu_{i_1}(\alloc_{i_1})} \times \frac{\valu_{i_2}(\alloc_{i_3})}{\valu_{i_2}(\alloc_{i_2})} \times \ldots \times  \frac{\valu_{i_k}(\alloc_{i})}{\valu_{i_k}(\alloc_{i_k})} & \\
		&= \frac{\valu(\alloc_{i_2})}{\valu(\alloc_{i_1})} \times \frac{\valu(\alloc_{i_3})}{\valu(\alloc_{i_2})} \times \ldots \times \frac{\valu(\alloc_{i})}{\valu(\alloc_{i_k})} &\text{Property \ref{prop42}} \\
		&= \frac{\valu(\alloc_i)}{\valu(\alloc_{i_1})}.
	\end{align*}
	
	Therefore $\virtualvalue_i(\alloc) = \frac{\valu(\alloc_i)}{\rank_i(\alloc_i)} = \valu(\alloc_i) \frac{\valu(\alloc_{i_1})}{\valu(\alloc_i)} = \valu(\alloc_{i_1}) = \valu(\alloc_{\roott{i}{\alloc}})$.	
\end{proof}

\begin{observationrep}\label{path-virtual}
	Let $\alloc$ be an allocation satisfying Property \ref{prop42}, and $i, j \in \agents$ be two agents that $i$ has a path to $j$ in $G_{0, \alloc}$. Then we have $\virtualvalue_j(\alloc) \le \virtualvalue_i(\alloc)$. 
\end{observationrep}
\begin{proof} 
	Let $r = \roott{i}{\alloc}$ and $r' = \roott{j}{\alloc}$. Since $r$ has a path to $j$, we have: 
	
	\begin{align*}
		\virtualvalue_j(\alloc) &= \valu(\alloc_{r'}) &\text{Observation \ref{obs41}} \\
		&\le \valu(\alloc_r) &\text{$r' = \roott{j}{\alloc}$} \\
		&= \virtualvalue_i(\alloc) &\text{Observation \ref{obs41}}. 
	\end{align*}
\end{proof}

\begin{lemmarep}\label{lem:prop43}
	Let $\alloc$ be an allocation satisfying Property \ref{prop42}, and suppose for every $i \in \agents$ and $\good \in \alloc_i$ we have $\valu(\alloc_i \setminus \{\good\}) \le \virtualvalue_i(\alloc)$. Then $\alloc$ satisfies Property \ref{prop43}. 
\end{lemmarep}
\begin{proof} 
	For every $j \in \agents$, we show that $\valu_j(\alloc_i \setminus \{\good\}) \le \virtualvalue_j(\alloc)$. Let $k$ be the corresponding agent of agent $i$. We have three cases $j = i$, $j = k$, and $j \notin \{i, k\}$. The first case is indeed the condition of the lemma. For the second case, we have: 
	
	\begin{align*}
		\valu_j(\alloc_i \setminus \{\good\}) &\le \valu(\alloc_i \setminus \{\good\}) &\\
		&\le \virtualvalue_i(\alloc) &\\
		&\le \virtualvalue_j(\alloc) &\text{Observation \ref{path-virtual}}\\
	\end{align*}
	
	And since $\alloc_i$ is relevant to only agents $i$ and $k$, for case $j \notin \{i, k\}$ we have $\valu_j(\alloc_i \setminus \{\good\}) = 0$ and the lemma holds trivially. 
\end{proof}

\subsection*{Update Rule 1}
This rule is applicable when an agent $i \in \agents$ with $\rank_i(\malloc) = 1$ and a good $\good \in \malloc_{0}$ exists such that 
	 $\good$ has the same type as the goods in $\malloc_i$, i.e., assuming $\malloc_i \subseteq \relevant{\goods}{i,j}$ for some $j$, then $\good \in \relevant{\goods}{i,j}$.
If so, we replace $\malloc_i$ with a minimal subset of $\malloc_i \cup \{\good\}$ that has a value more than $\valu(\malloc_i)$. 

\begin{observation}\label{graph-unchanged}
Let $\malloc'$ be the result of applying Rule 1 to $\malloc$. Then, all edges in $G_{0, \malloc}$ remain unchanged in $G_{0, \malloc'}$.
\end{observation}

\begin{lemmarep}\label{lem:U81}
	Let $\malloc$ be an allocation satisfying Properties \ref{prop42} and \ref{prop43}, and let $\malloc'$ be the result of applying Rule 1 on $\malloc$. Then, $\malloc'$ satisfies Properties \ref{prop42} and \ref{prop43}. Furthermore, we have $\phi(\malloc) < \phi(\malloc')$ and for every remaining agent $k$ we have $\virtualvalue_k(\malloc)\leq \virtualvalue_k(\malloc')$.
\end{lemmarep}
\begin{proof} 
	Given that \(\malloc'_i \subseteq \relevant{\goods}{i,j}\) and the other bundles remain unchanged, \(\malloc'\) satisfies Property \ref{prop42}.
	By Lemma \ref{lem:prop43}, to prove that $\malloc'$ satisfies Property \ref{prop43}, it's enough to show that for every remaining agent $k$ and $\good \in \malloc'_k$, we have $\valu(\malloc'_k \setminus \{\good\}) \le \virtualvalue_k(\malloc')$. There are two cases $k = i$ and $k \neq i$.
	Let \(k \neq i\) and a good \(\good \in \malloc'_k\) and \(r = \roott{k}{\malloc'}\). We have:
	
	\begin{align*}
		\valu(\malloc'_k \setminus \{\good\}) &= \valu(\malloc_k \setminus \{\good\}) & \text{$k \neq i$} \\
		&\le \virtualvalue_k(\malloc) &\text{Property \ref{prop43}} \\
		& \le \valu(\malloc_{r}) & \text{Observation \ref{graph-unchanged}} \\
		& \le \valu(\malloc'_{r}) & \\
		& = \virtualvalue_k(\malloc') & \text{Observation \ref{obs41}}.
	\end{align*}
	
	Now let $k = i$ and a good \(\good \in \malloc'_i\) and \(r = \roott{i}{\malloc'}\). We have:
	
	\begin{align*}
		\valu(\malloc'_i \setminus \{\good\}) & \le \valu(\malloc_i) & \text{\(\malloc'_i\) is minimal} \\
		& \le \valu(\malloc_{r}) & \text{$\rank_i(\malloc) = 1$} \\
		& \le \valu(\malloc'_{r}) & \\
		& = \virtualvalue_i(\malloc') & \text{Observation \ref{obs41}}.
	\end{align*}
	
	Therefore, \(\malloc'\) satisfies both Properties \ref{prop42} and \ref{prop43}.
	
	Since $\valu(\malloc_i) < \valu(\malloc'_i)$, and all other bundles remain unchanged, we have $\phi(\malloc) < \phi(\malloc')$. 
	
	In order to show that for every agent $k$ we have $\virtualvalue_k(\malloc) \le \virtualvalue_k(\malloc')$, let $r = \roott{k}{\malloc'}$. We have: 
	
	\begin{align*}
		\virtualvalue_k(\malloc) &\le \valu(\malloc_r) &\text{Observation \ref{graph-unchanged}} \\
		&\le \valu(\malloc'_r) & \\
		&= \virtualvalue_k(\malloc') &\text{Observation \ref{obs41}}.  
	\end{align*}
\end{proof}

\subsection*{Update Rule 2}
This rule is applicable if there exists an agent $i \in \agents$ such that for some $j$, $\virtualvalue_i(\malloc)< \valu(\relevant{\malloc_{0}}{i,j}).$ If so, we update the allocation as follows: let $j$ be the agent with the minimum virtual value such that $\virtualvalue_i(\malloc)< \valu(\relevant{\malloc_{0}}{i,j})$, and let $P = i_1 \rightarrow i_2 \rightarrow \ldots \rightarrow i_k \rightarrow i$ be the rankpath of agent $i$. We allocate a minimal subset of $\relevant{\malloc_0}{i,j}$ with value more than $\virtualvalue_i(\malloc)$ to agent $i$, shift the bundles through $P$, and return $\malloc_{i_1}$ back to the pool.

\begin{lemmarep}\label{lem:U82}
	Let $\malloc$ be an allocation satisfying Properties \ref{prop42} and\ref{prop43}, and let $\malloc'$ be the result of applying Rule 2 on $\malloc$. Then, $\malloc'$ satisfies Properties \ref{prop42} and \ref{prop43}. Furthermore, we have $\phi(\malloc) < \phi(\malloc')$ and for every remaining agent $k$ we have $\virtualvalue_k(\malloc) \leq \virtualvalue_k(\malloc')$.
\end{lemmarep}
\begin{proof}
	During the update, only the bundle of agent $i$ is modified. \footnote{While other bundles may be transferred among agents, none is modified.} Therefore, Property \ref{prop42} remains valid for agents other than $i$. For agent $i$ we also know that $\malloc'_i \subseteq \relevant{\goods}{i, j}$. Therefore, Property \ref{prop42} holds for $\malloc'$. 
	
	To show that Property \ref{prop43} holds after the update, we note that this update rule is indeed exactly the same as Rule 1 in Section \ref{sec:eh}. Therefore, by Lemma \ref{virtual-value-increases} we conclude that virtual values do not decrease after the update. Furthermore, by Lemma \ref{no-virtual-se}, $\malloc'$ satisfies Property \ref{prop43}, and by Lemma \ref{nsw-increases} we have $\phi(\malloc) < \phi(\malloc')$.
\end{proof}

\subsection*{Update Rule 3}
This rule is applicable when there is an agent $i \in \agents$ such that for every agent $j \neq i$, $\valu_i(\malloc_j) = 0$. If so, we ask agent $i$ to pick either $\malloc_i$ or $\relevant{\malloc_0}{i}$ that is more valuable to him and return the other bundle back to the pool.
Also, we mark agent $i$ as satisfied and remove her from the process.

\begin{lemmarep}\label{lem:U83}
	Let $\malloc$ be an allocation satisfying Properties \ref{prop42} and \ref{prop43}, and let $\malloc'$ be the result of applying Rule 3 on $\malloc$. Then, $\malloc'$ satisfies Properties \ref{prop42} and \ref{prop43}. Furthermore, we have $\phi(\malloc) < \phi(\malloc')$ and for every remaining agent $k$ we have $\virtualvalue_k(\malloc) \leq \virtualvalue_k(\malloc')$.
\end{lemmarep}
\begin{proof} 
	According to Rule 3, for every remaining agent $j$, we have \( \valu_i(\malloc_j) = 0 \), meaning agent \( i \) isn't in agent \( j \)'s rankpath. Thus, \( \virtualvalue_j(\malloc) = \virtualvalue_j(\malloc') \).
	Also, the update keeps all the remaining agents' bundles unchanged. Thus, \( \malloc' \) satisfies Properties \ref{prop42} and \ref{prop43}.
	Finally, since $|\agents|$ decreases, we have \( \phi(\malloc) < \phi(\malloc') \).
\end{proof}

\begin{lemmarep}\label{no-se-1}
	Let $i$ be an agent satisfied by Rule 3. Then, $i$ will never strongly envy any remaining agent, and no remaining agent will ever strongly envy agent $i$.
\end{lemmarep}
\begin{proof} 
	After the update, we have $\valu_i(\malloc'_0) \le \valu_i(\malloc^*_i)$, and for every remaining agent $j$, $\valu_i(\malloc'_j) = 0$. Therefore, $\valu_i(\cup \malloc') \le \valu_i(\malloc^*_i)$ which means that agent $i$ will not envy any remaining agent in subsequent steps.
	Furthermore, considering Property \ref{prop42} for $\malloc$, there exists an agent $j$ such that $\malloc_i \subseteq \relevant{\goods}{i, j}$. The only agent who might strongly envy $i$ is agent $j$. Post-update, $\malloc'_i$ can either be $\malloc_i$ or $\relevant{\malloc_0}{i}$.
	
	If $\malloc'_i = \malloc_i$, let $\good \in \malloc'_i$, and let $\malloc^*$ denote the final allocation. We have:
	
	\[
	\begin{aligned}
		\valu(\malloc'_i \setminus \{\good\}) &= \valu(\malloc_i \setminus \{\good\}) & \text{$\malloc'_i = \malloc_i$} \\
		&\le \virtualvalue_i(\malloc) & \text{Property \ref{prop43}} \\
		&\le \virtualvalue_j(\malloc) & \text{Observation \ref{path-virtual}} \\
		&\le \virtualvalue_j(\malloc^*) &  \\
		&\le \valu(\malloc^*_j).
	\end{aligned}
	\]
	
	Thus, agent $j$ does not strongly envy $i$ in the final allocation.
	
	Now, if $\malloc'_i = \relevant{\malloc_0}{i}$, for any remaining agent $j \in \agents$ in the final allocation $\malloc^*$, we have:
	
	\[
	\begin{aligned}
		\valu_j(\malloc'_i) &= \valu(\relevant{\malloc_0}{i, j}) & \text{$\malloc'_i = \relevant{\malloc_0}{i}$} \\
		&\le \virtualvalue_j(\malloc) & \text{Rule 2 is not applicable} \\
		&\le \virtualvalue_j(\malloc^*) & \\
		&\le \valu(\malloc^*_j).
	\end{aligned}
	\]
	
	Therefore, agent $j$ does not strongly envy $i$ in the final allocation.
\end{proof}

\subsection*{Update Rule 4}
First, we prove that when the first three rules are not applicable, then $G_{0, \malloc}$ has a special structure. 

\begin{lemmarep}\label{lem:cycle}
	Let $\malloc$ be an allocation satisfying Properties \ref{prop42} and \ref{prop43}. If the first three rules are not applicable, then  $G_{0, \malloc}$ consists of one or several disjoint cycles.
\end{lemmarep}
\begin{proof}
	Since $\malloc$ satisfies Property \ref{prop42}, the in-degree of each vertex in $G_{0, \malloc}$ is equal to $1$. Additionally since Rule 3 is not applicable to $\malloc$, the out-degree of each vertex is at least $1$. Therefore $G_{0, \malloc}$ consists of one or more cycles. 
\end{proof}

In the fourth updating rule, we select and satisfy one agent with rank 1. Let $i$ be an agent with $\rank_i(\malloc)=1$ and suppose $i \rightarrow j_1 \rightarrow j_2 \ldots \rightarrow j_k \rightarrow i$ be the cycle in $G_{0, \malloc}$ containing $i$. We ask agent $i$ to pick either $\malloc_i \cup \relevant {\malloc_{0}}{i}$ or $\malloc_{j_1}$, whichever that is more valuable to her. We then mark agent $i$ as satisfied and remove her from the process. Finally, if agent $i$ picked $\malloc_{j_1}$, we shift the bundles through the cycle, i.e., for every $ 2 \leq \ell<k$, we give $\malloc_{\ell+1}$ to agent $j_{\ell}$ and give $\malloc_i$ to agent $j_k$.

\begin{lemmarep}\label{lem:U84}
	Let $\malloc$ be an allocation satisfying Properties \ref{prop42} and \ref{prop43} and let $\malloc'$ be the result of applying Rule 4 on $\malloc$. Then, $\malloc'$ satisfies Properties \ref{prop42} and \ref{prop43}. Furthermore, we have $\phi(\malloc) < \phi(\malloc')$ and for every remaining agent $k$ we have $\virtualvalue_k(\malloc) \leq \virtualvalue_k(\malloc')$.
\end{lemmarep}
\begin{proof} 
	First we show that the virtual value of each remaining agent does not decrease after the update. 	
	By Observation \ref{path-virtual} we have $\virtualvalue_i(\malloc) = \virtualvalue_{j_1}(\malloc) = \virtualvalue_{j_2}(\malloc) = \ldots = \virtualvalue_{j_k}(\malloc) = \valu(\malloc_i)$.
	Now let $r' = \roott{k}{\malloc'}$ and let $l$ be the agent such that $\malloc'_{r'} = \malloc_l$. For every remaining agent $j$ we have: 
	\begin{align*}
		\virtualvalue_j(\malloc) &= \virtualvalue_l(\malloc) & \\
		&\le \valu(\malloc_{l}) & \\
		&= \valu(\malloc'_{r'}) &\text{$\malloc'_{r'} = \malloc_l$} \\
		&= \virtualvalue_j(\malloc') &\text{$\roott{k}{\malloc'} = r'$}. 
	\end{align*}
	Hence, the virtual values of the remaining agents do not decrease.
	
	Since the set of bundles in $\malloc'$ remains unchanged, 	$\malloc'$ satisfies Property \ref{prop42}. 
	To Prove that $\malloc'$ satisfies Property \ref{prop43}, let $j$ be the remaining agent and $\good \in \malloc'_j$ and $\malloc'_j = \malloc_k$. We have: 
	
	\begin{align*}
		\valu(\malloc'_j \setminus \{\good\}) &= \valu(\malloc_k \setminus \{\good\}) &\text{$\malloc'_j = \malloc_k$} \\
		&\le \virtualvalue_k(\malloc) &\text{Property \ref{prop43}} \\
		&= \virtualvalue_j(\malloc) & \\
		&\le \virtualvalue_j(\malloc').
	\end{align*}
	
	Therefore $\malloc'$ satisfies both Properties \ref{prop42} and \ref{prop43}. 
	
	Since $|\agents'| = |\agents| - 1$, we have $\phi(\malloc) < \phi(\malloc')$. 
\end{proof}

\begin{lemmarep}\label{no-se-2}
	Let $i$ be an agent satisfied by Rule 4. Then, $i$ will never strongly envy any remaining agent, and no remaining agent will ever strongly envy agent $i$.
\end{lemmarep}
\begin{proof} 
	Since only $\malloc_0, \malloc_i, \malloc_{j_1}$ may be relevant to agent $i$, and this agent picks the more valuable bundle between $\malloc_i \cup \relevant{\malloc_0}{i}$ and $\malloc_{j_1}$, it doesn't matter what happens in the future, and $i$ will not envy any bundle in $\malloc^*$. 
	
	Suppose that $j$ is the corresponding agent of agent $i$. 
	There are two cases for $\malloc^*_i$. 
	The first case is $\malloc'_i = \malloc_i \cup  \relevant{\malloc_0}{i}$. In this case for agent $k \neq j$ we have: 
	
	\begin{align*}
		\valu_k(\malloc^*_i) &= \relevant{\malloc_0}{i, k}& \text{$\malloc_i \subseteq \relevant{\goods}{i, j}$} \\
		&\le \virtualvalue_k(\malloc) &\text{Rule 2 is not applicable} \\
		&\le \virtualvalue_k(\malloc^*) & \\
		&\le \valu_k(\malloc_k^*). 
	\end{align*}
	
	and for agent $j$ we have: 
	
	\begin{align*}
		\valu_j(\malloc^*_i) &= \valu(\relevant{\malloc_i}{i, j}) &\text{Rule 1 is not applicable} \\
		&= \valu(\malloc_i) &\text{$\malloc_i \subseteq \relevant{\goods}{i, j}$} \\
		&= \virtualvalue_j(\malloc) &\text{Observation \ref{obs41}} \\
		&\le \virtualvalue_j(\malloc^*) & \\
		&\le \valu_j(\malloc_j^*). 
	\end{align*}
	
	The second case is $\malloc'_i = \malloc_{j_1}$. In this case only $j_1$ may envy $i$. Let $\good \in \malloc_{j_1}$. We have: 
	
	\begin{align*}
		\valu(\malloc^*_i \setminus \{\good\}) &= \valu(\malloc_{j_1} \setminus \{\good\}) &\text{$\malloc^*_i = \malloc_{j_1}$} \\
		&\le \virtualvalue_{j_1}(\malloc) &\text{Property \ref{prop43}}
		\\
		&\le \virtualvalue_{j_1}(\malloc^*) &\\
		&\le \valu_{j_1}(\malloc_{j_1}^*). 
	\end{align*}
	
	Therefore no agent in $\malloc^*$ will strongly envy $i$.  
\end{proof}

\subsection*{Final Step}
Finally, if only one agent is left behind, i.e., $|\agents| = 1$, we allocate the whole set of items in $\malloc_{0}$ to the remaining agent and return the allocation. 

\begin{theoremrep}\label{thm8}
	Suppose the valuations are restricted additive and $(2, \infty)$-bounded, and let $\malloc^*$ be the allocation returned by $\algopr$. Then, $\malloc^*$ is an $\efx$ allocation. 
\end{theoremrep}
\begin{proof}
	In this algorithm, we satisfy all agents one by one. By Lemmas \ref{no-se-1} and \ref{no-se-2}, the returned allocation by $\algopr$ is $\efx$. 
\end{proof}

	\newpage
	\bibliographystyle{abbrv}
	\bibliography{draft}

\begin{thebibliography}{10}

\bibitem{akrami2022efx}
H.~Akrami, N.~Alon, B.~R. Chaudhury, J.~Garg, K.~Mehlhorn, and R.~Mehta.
\newblock Efx allocations: Simplifications and improvements.
\newblock {\em arXiv preprint arXiv:2205.07638}, 2022.

\bibitem{akrami2023efx}
H.~Akrami, N.~Alon, B.~R. Chaudhury, J.~Garg, K.~Mehlhorn, and R.~Mehta.
\newblock Efx: a simpler approach and an (almost) optimal guarantee via rainbow
  cycle number.
\newblock In {\em Proceedings of the 24th ACM Conference on Economics and
  Computation}, pages 61--61, 2023.

\bibitem{akrami2022ef2x}
H.~Akrami, R.~Rezvan, and M.~Seddighin.
\newblock An ef2x allocation protocol for restricted additive valuations.
\newblock {\em arXiv preprint arXiv:2202.13676}, 2022.

\bibitem{amanatidis2022fair}
G.~Amanatidis, G.~Birmpas, A.~Filos-Ratsikas, and A.~A. Voudouris.
\newblock Fair division of indivisible goods: A survey.
\newblock {\em arXiv preprint arXiv:2202.07551}, 2022.

\bibitem{amanatidis2024pushing}
G.~Amanatidis, A.~Filos-Ratsikas, and A.~Sgouritsa.
\newblock Pushing the frontier on approximate efx allocations.
\newblock {\em arXiv preprint arXiv:2406.12413}, 2024.

\bibitem{amanatidis2020multiple}
G.~Amanatidis, E.~Markakis, and A.~Ntokos.
\newblock Multiple birds with one stone: Beating 1/2 for efx and gmms via envy
  cycle elimination.
\newblock {\em Theoretical Computer Science}, 841:94--109, 2020.

\bibitem{annamalai2017combinatorial}
C.~Annamalai, C.~Kalaitzis, and O.~Svensson.
\newblock Combinatorial algorithm for restricted max-min fair allocation.
\newblock {\em ACM Transactions on Algorithms (TALG)}, 13(3):1--28, 2017.

\bibitem{asadpour2012santa}
A.~Asadpour, U.~Feige, and A.~Saberi.
\newblock Santa claus meets hypergraph matchings.
\newblock {\em ACM Transactions on Algorithms (TALG)}, 8(3):1--9, 2012.

\bibitem{aziz2022algorithmic}
H.~Aziz, B.~Li, H.~Moulin, and X.~Wu.
\newblock Algorithmic fair allocation of indivisible items: A survey and new
  questions.
\newblock {\em ACM SIGecom Exchanges}, 20(1):24--40, 2022.

\bibitem{aziz2016discrete}
H.~Aziz and S.~Mackenzie.
\newblock A discrete and bounded envy-free cake cutting protocol for any number
  of agents.
\newblock In {\em Foundations of Computer Science (FOCS), 2016 IEEE 57th Annual
  Symposium on}, pages 416--427. IEEE, 2016.

\bibitem{bamas2023better}
{\'E}.~Bamas and L.~Rohwedder.
\newblock Better trees for santa claus.
\newblock In {\em Proceedings of the 55th Annual ACM Symposium on Theory of
  Computing}, pages 1862--1875, 2023.

\bibitem{barman2023parameterized}
S.~Barman, D.~Kar, and S.~Pathak.
\newblock Parameterized guarantees for almost envy-free allocations.
\newblock {\em arXiv preprint arXiv:2312.13791}, 2023.

\bibitem{barman2017finding}
S.~Barman, S.~K. Krishnamurthy, and R.~Vaish.
\newblock Finding fair and efficient allocations.
\newblock In {\em Proceedings of the 2018 ACM Conference on Economics and
  Computation}, pages 557--574, 2018.

\bibitem{berendsohn2022fixed}
B.~A. Berendsohn, S.~Boyadzhiyska, and L.~Kozma.
\newblock Fixed-point cycles and efx allocations.
\newblock {\em arXiv preprint arXiv:2201.08753}, 2022.

\bibitem{berger2022almost}
B.~Berger, A.~Cohen, M.~Feldman, and A.~Fiat.
\newblock Almost full efx exists for four agents.
\newblock In {\em Proceedings of the AAAI Conference on Artificial
  Intelligence}, volume~36, pages 4826--4833, 2022.

\bibitem{Bezacova:first}
I.~Bez{\'a}kov{\'a} and V.~Dani.
\newblock Allocating indivisible goods.
\newblock {\em ACM SIGecom Exchanges}, 5(3):11--18, 2005.

\bibitem{brams}
S.~J. Brams and A.~D. Taylor.
\newblock An envy-free cake division protocol.
\newblock {\em American Mathematical Monthly}, pages 9--18, 1995.

\bibitem{brams1996fair}
S.~J. Brams and A.~D. Taylor.
\newblock {\em Fair Division: From cake-cutting to dispute resolution}.
\newblock Cambridge University Press, 1996.

\bibitem{Budish:first}
E.~Budish.
\newblock The combinatorial assignment problem: Approximate competitive
  equilibrium from equal incomes.
\newblock {\em Journal of Political Economy}, 119(6):1061--1103, 2011.

\bibitem{caragiannis2022existence}
I.~Caragiannis, J.~Garg, N.~Rathi, E.~Sharma, and G.~Varricchio.
\newblock Existence and computation of epistemic efx allocations.
\newblock {\em arXiv e-prints}, pages arXiv--2206, 2022.

\bibitem{caragiannis2019envy}
I.~Caragiannis, N.~Gravin, and X.~Huang.
\newblock Envy-freeness up to any item with high nash welfare: The virtue of
  donating items.
\newblock In {\em Proceedings of the 2019 ACM Conference on Economics and
  Computation}, pages 527--545, 2019.

\bibitem{caragiannis2016unreasonable}
I.~Caragiannis, D.~Kurokawa, H.~Moulin, A.~D. Procaccia, N.~Shah, and J.~Wang.
\newblock The unreasonable fairness of maximum nash welfare.
\newblock In {\em Proceedings of the 2016 ACM Conference on Economics and
  Computation}, pages 305--322. ACM, 2016.

\bibitem{caragiannis2019unreasonable}
I.~Caragiannis, D.~Kurokawa, H.~Moulin, A.~D. Procaccia, N.~Shah, and J.~Wang.
\newblock The unreasonable fairness of maximum nash welfare.
\newblock {\em ACM Transactions on Economics and Computation (TEAC)},
  7(3):1--32, 2019.

\bibitem{chaudhury2020efx}
B.~R. Chaudhury, J.~Garg, and K.~Mehlhorn.
\newblock Efx exists for three agents.
\newblock In {\em Proceedings of the 21st ACM Conference on Economics and
  Computation}, pages 1--19, 2020.

\bibitem{chaudhury2021improving}
B.~R. Chaudhury, J.~Garg, K.~Mehlhorn, R.~Mehta, and P.~Misra.
\newblock Improving efx guarantees through rainbow cycle number.
\newblock In {\em Proceedings of the 22nd ACM Conference on Economics and
  Computation}, pages 310--311, 2021.

\bibitem{chaudhury2021little}
B.~R. Chaudhury, T.~Kavitha, K.~Mehlhorn, and A.~Sgouritsa.
\newblock A little charity guarantees almost envy-freeness.
\newblock {\em SIAM Journal on Computing}, 50(4):1336--1358, 2021.

\bibitem{chen2009settling}
X.~Chen, X.~Deng, and S.-H. Teng.
\newblock Settling the complexity of computing two-player nash equilibria.
\newblock {\em Journal of the ACM (JACM)}, 56(3):1--57, 2009.

\bibitem{christodoulou2023fair}
G.~Christodoulou, A.~Fiat, E.~Koutsoupias, and A.~Sgouritsa.
\newblock Fair allocation in graphs.
\newblock In {\em Proceedings of the 24th ACM Conference on Economics and
  Computation}, pages 473--488, 2023.

\bibitem{daskalakis2009complexity}
C.~Daskalakis, P.~W. Goldberg, and C.~H. Papadimitriou.
\newblock The complexity of computing a nash equilibrium.
\newblock {\em Communications of the ACM}, 52(2):89--97, 2009.

\bibitem{davies2020tale}
S.~Davies, T.~Rothvoss, and Y.~Zhang.
\newblock A tale of santa claus, hypergraphs and matroids.
\newblock In {\em Proceedings of the Fourteenth Annual ACM-SIAM Symposium on
  Discrete Algorithms}, pages 2748--2757. SIAM, 2020.

\bibitem{Dubins:first}
L.~E. Dubins and E.~H. Spanier.
\newblock How to cut a cake fairly.
\newblock {\em American mathematical monthly}, pages 1--17, 1961.

\bibitem{even1984note}
S.~Even and A.~Paz.
\newblock A note on cake cutting.
\newblock {\em Discrete Applied Mathematics}, 7(3):285--296, 1984.

\bibitem{farhadi2021almost}
A.~Farhadi, M.~Hajiaghayi, M.~Latifian, M.~Seddighin, and H.~Yami.
\newblock Almost envy-freeness, envy-rank, and nash social welfare matchings.
\newblock In {\em Proceedings of the AAAI Conference on Artificial
  Intelligence}, volume~35, pages 5355--5362, 2021.

\bibitem{ghodsi2018fair}
M.~Ghodsi, M.~HajiAghayi, M.~Seddighin, S.~Seddighin, and H.~Yami.
\newblock Fair allocation of indivisible goods: Improvements and
  generalizations.
\newblock In {\em Proceedings of the 2018 ACM Conference on Economics and
  Computation}, pages 539--556. ACM, 2018.

\bibitem{gourves2014near}
L.~Gourv{\`e}s, J.~Monnot, and L.~Tlilane.
\newblock Near fairness in matroids.
\newblock In {\em ECAI}, volume~14, pages 393--398, 2014.

\bibitem{hall1987representatives}
P.~Hall.
\newblock On representatives of subsets.
\newblock {\em Classic Papers in Combinatorics}, pages 58--62, 1987.

\bibitem{jahan2022rainbow}
S.~C. Jahan, M.~Seddighin, S.-M. Seyed-Javadi, and M.~Sharifi.
\newblock Rainbow cycle number and efx allocations:(almost) closing the gap.
\newblock {\em arXiv preprint arXiv:2212.09482}, 2022.

\bibitem{khot2007approximation}
S.~Khot and A.~K. Ponnuswami.
\newblock Approximation algorithms for the max-min allocation problem.
\newblock In {\em International Workshop on Approximation Algorithms for
  Combinatorial Optimization}, pages 204--217. Springer, 2007.

\bibitem{Procaccia:first}
D.~Kurokawa, A.~D. Procaccia, and J.~Wang.
\newblock Fair enough: Guaranteeing approximate maximin shares.
\newblock {\em Journal of the ACM (JACM)}, 65(2):8, 2018.

\bibitem{Saberi:first}
R.~J. Lipton, E.~Markakis, E.~Mossel, and A.~Saberi.
\newblock On approximately fair allocations of indivisible goods.
\newblock In {\em Proceedings of the 5th ACM conference on Electronic
  commerce}, pages 125--131. ACM, 2004.

\bibitem{plaut2018almost}
B.~Plaut and T.~Roughgarde.
\newblock Almost envy-freeness with general valuations.
\newblock In {\em Proceedings of the Twenty-Ninth Annual ACM-SIAM Symposium on
  Discrete Algorithms}, pages 2584--2603. SIAM, 2018.

\bibitem{procaccia2013cake}
A.~D. Procaccia.
\newblock Cake cutting: Not just child's play.
\newblock {\em Communications of the ACM}, 56(7):78--87, 2013.

\bibitem{procaccia2020technical}
A.~D. Procaccia.
\newblock Technical perspective: An answer to fair division's most enigmatic
  question.
\newblock {\em Communications of the ACM}, 63(4):118--118, 2020.

\bibitem{Steinhaus:first}
H.~Steinhaus.
\newblock The problem of fair division.
\newblock {\em Econometrica}, 16(1), 1948.

\bibitem{stromquist1980cut}
W.~Stromquist.
\newblock How to cut a cake fairly.
\newblock {\em The American Mathematical Monthly}, 87(8):640--644, 1980.

\end{thebibliography}
	\appendix
	\newpage
	\appendix
	\section{Missing Proofs}

Lemma \ref{lem:phi} helps us prove some of our claims in Section \ref{sec:5}. A Lemma similar to Lemma \ref{lem:phi} has been previously proved by Akrami \etal.\cite{akrami2022ef2x}. For completeness, here we restate and prove this lemma.
\begin{lemma}\label{lem:phi}
	Let $\alloc$ and $\alloc'$ be two allocations. Define:
	\begin{align*}
		A = \{i | \alloc_i \neq \alloc'_i\} \\
		x = \min_{i \in A}(\valu(\alloc_i)) \\
		x' = \min_{i \in A}(\valu(\alloc'_i)).
	\end{align*}
	Assume the following conditions hold:
	\begin{itemize}
		\item $x < x'$,
		\item There is only one source agent in $A$ in $G{\beta,\alloc}$,
		\item If $i$ is the source agent in $A$, then $\valu(\alloc_i)=x$.
	\end{itemize}
	Then, $\phi(\alloc) < \phi(\alloc')$.
\end{lemma}
\begin{proof}
	Denote the sources of $G_{\beta, \alloc}$ by $s_1, s_2, \ldots, s_k$ and the sources of $G_{\beta, \alloc'}$ by $s'_1, s'_2, \ldots, s'_{k'}$. Then:
	\begin{align*}
		\phi(\alloc) = \big[ \valu(\alloc_{s_1}),\valu(\alloc_{s_2}),\ldots, \valu(\alloc_{s_k}),\infty \big], \\
		\phi(\alloc') = \big[ \valu(\alloc'_{s'_1}),\valu(\alloc'_{s'_2}),\ldots, \valu(\alloc'_{s'_{k'}}),\infty \big].
	\end{align*}
	Since agent $i$ is a source in $G_{\beta, \alloc}$, we can assume without loss of generality that there exist an index $l$ such that $s_l = i$, and $\valu(\alloc_{s_l}) \neq \valu(\alloc{s_{l+1}})$. (we can reorder same values in $\phi$ to satisfy this condition). We show that for every $l' < l$ we have $\phi(\alloc)_{l'} = \phi(\alloc')_{l'}$ and for $l$ we have $\phi(\alloc)_l < \phi(\alloc')_l$. 
	\begin{itemize}
		\item For every $l' < l$, we claim that $s_{l'}$ remains a source in $G_{\beta, \alloc'}$. To prove this, we show that for every agent $j$ we have $\valu_j(\alloc'_{s_{l'}}) < \beta \valu_j(\alloc'_j)$. If $j \notin A$, our claimi is trivial since $s_{l'}$ is a source in $G_{\beta, \alloc}$. If $j \in A$, we have: 
		\begin{align*}
			\valu_j(\alloc'_{s_{l'}}) &\le \valu(\alloc'_{s_{l'}}) &\\
			&= \valu(\alloc_{s_{l'}}) &\text{$s_{l'} \notin A$}\\
			&\le x &\text{$l' < l$}, \\
			&< x' \\
			&\le \valu(\alloc'_j) &\text{$j \in A$},\\
			&= \valu_j(\alloc'_j) &\\
			&\le \beta \valu_j(\alloc'_j). 
		\end{align*}
		Therefore $j$ does not $\beta$-envy $s_{l'}$ in $\alloc'$ and $\phi(\alloc)_{l'} = \phi(\alloc')_{l'}$. 
		
		\item To prove $\phi(\alloc)_l < \phi(\alloc')_l$, we need to show that for any source agent $j$ in $G_{\beta, \alloc'}$ 
		such that $j \notin \{s_1, s_2, \dots, s_{l - 1}\}$ we have $\valu(\alloc'_j) > x = \phi(\alloc)_l$. 
		We have three cases. 
		The first case is $j \in A$. In this case we have: 
		\begin{align*}
			\phi(\alloc)_l &= x &\\
			& < x'\\
			&\le \valu(\alloc'_j) &\text{$j \in A$}. 
		\end{align*}
		In the second case we have $j \notin A$ and $j$ is also a source in $G_{\beta, \alloc}$. In this case we have: 
		\begin{align*}
			\phi(\alloc)_l &= x &\\
			&< \valu(\alloc_j) &\text{$j \notin {s_1, s_2, \dots, s_{l - 1}}$},\\
			&= \valu(\alloc'_j) &\text{$j \notin A$}. 
		\end{align*}
		In the last case we have $j \notin A$ and $j$ is not a source in $G_{\beta, \alloc}$. We know there exists another agent $j' \in A$ such that $j'$ $\beta$-envies agent $j$ in $G_{\beta, \alloc}$. Therefore we have: 
		\begin{align*}
			\phi(\alloc)_l &= x &\\
			&< \sqrt{2} x &\\
			& \le\sqrt{2} \valu(\alloc_{j'}) &\text{$j' \in A$},\\
			&< \valu(\alloc_j) &\text{$j'$ envies $j$ in $\alloc$},\\
			&= \valu(\alloc'_j) &\text{$j \notin A$}. 
		\end{align*}
	\end{itemize}
	Hence, $\phi(\alloc')_l$ is replaced with an agent with a larger value, or $\phi(\alloc')_l = \infty$. 
	
\end{proof}

\end{document}